\documentclass[aps,pra,twocolumn,superscriptaddress]{revtex4}
\usepackage{graphicx}
\usepackage{amsmath}
\usepackage{amsfonts}%
\usepackage{amssymb}
\usepackage{mathrsfs}
\usepackage{dcolumn}
\usepackage{bm}
\usepackage{float}
\usepackage{color}
\usepackage{epsfig}

\begin{document}

\title{Multicore fibre photonic lanterns for precision radial velocity science}

\author{Itandehui Gris-S\'{a}nchez	} 
\email[]{itan.gs@gmail.com}
\affiliation{Centre for Photonics and Photonic Materials, Department of Physics, University of Bath, Bath, BA2 7AY, UK}
\author{Dionne M. Haynes}
\affiliation{Leibniz-Institut für Astrophysik Potsdam, An der Sternwarte 16, 14482 Potsdam, Germany}
\affiliation{innoFSPEC Potsdam, An der Sternwarte 16, 14482 Potsdam, Germany}
\author{Katjana Ehrlich}
\affiliation{Leibniz-Institut für Astrophysik Potsdam, An der Sternwarte 16, 14482 Potsdam, Germany}
\author{Roger Haynes}
\affiliation{Leibniz-Institut für Astrophysik Potsdam, An der Sternwarte 16, 14482 Potsdam, Germany}
\affiliation{innoFSPEC Potsdam, An der Sternwarte 16, 14482 Potsdam, Germany}
\author{Tim A. Birks}
\affiliation{Centre for Photonics and Photonic Materials, Department of Physics, University of Bath, Bath, BA2 7AY, UK}

\begin{abstract}
Envisioning more compact and cost accessible astronomical instruments is now possible with existing photonic technologies like specialty optical fibres, photonic lanterns and ultrafast laser inscribed chips. We present an original design of a multicore fibre (MCF) terminated with multimode photonic lantern ports. It is designed to act as a relay fibre with the coupling efficiency of a multimode fibre, modal stability similar to a single-mode fibre and low loss in a wide range of wavelengths (380~nm to 860~nm). It provides phase and amplitude scrambling to achieve a stable near field and far field output illumination pattern despite input coupling variations, and low modal noise for increased photometric stability for high signal-to-noise applications such as precision radial velocity (PRV) science. Preliminary results are presented for a 511-core MCF and compared with current state of the art octagonal fibre.
\end{abstract}

\maketitle

\section{Introduction.}

Much effort and resources is devoted to control the mechanical and thermal stability of complex optical instruments like those used in astronomy. The construction and maintenance of increasingly sophisticated instruments is expensive, and they often require convoluted solutions to achieve the performance required. To extract information from all the useful photons gathered by sophisticated telescopes remains a challenge. Modal noise and focal ratio degradation (FRD), for example, degrade a spectrograph's performance and ultimately limit the ability to draw definite conclusions from an observation.

Optical fibres are frequently used in astronomical instruments, for example for interferometry or as part of multiobject imaging systems\cite{Heacox1992,Parry1998,BlandHawthorn2009,BlandHawthorn2012}. Incomplete scrambling of the fibre modal noise impacts high spectral resolution and high signal-to-noise applications, which limits the precision of measurements. During an astronomical observation the light coupling into the fibre can change due to varying seeing conditions, telescope guiding inaccuracies and beam defocus. This along with perturbation of the fibre along its length causes the light pattern at the output of a multimode fibre to change, which introduces apparent spectral line shifts, variable spectral line widths, and photometric inaccuracies (that limit signal-to-noise ratios) due to fibre modal noise \cite{Baudrand2001,Feger2012,Grupp2003,Lemke2011,McCoy2012,Origlia2014} which cannot all be completely calibrated out. 

Multicore  fibres (MCF's) and photonic lanterns (PL's) \cite{Birks2015} are compact photonic technologies that require less elaborated setups to incorporate them to other systems and which need little maintenance. Used as relays between telescope and instruments, they can also reduce modal noise or filter out narrow absorption lines \cite{Trinh2013,BlandHawthorn2004,Baudrand2001}. Their versatility therefore makes them good candidates for implementation in astronomical instruments. Progress has already been made for the improvement of current instruments capabilities by incorporating photonic technologies\cite{Content2014,Trinh2013,BlandHawthorn2012,Halverson2015,Jovanovic2016,MacLachlan2016}. Although astronomical instruments under current development continue to consider the use of octagonal and rectangular fibres for modal noise scrambling \cite{Shuterland2016}, they could benefit from incorporating custom-made photonic devices in such instruments as that proposed in this paper.

Birks \cite{Birks2012} observed modal noise scrambling in a device comprising an MCF (originally intended for the filtering of the narrow OH emission lines using fibre Bragg gratings) with 121 identical cores terminated with multimode (MM) photonic lantern ports \cite{Birks2011,Birks2015,LeonSaval2010,Olaya2012} that could be retrofitted to an optical high-resolution facility whilst maintaining efficient telescope to spectrograph coupling. Nonetheless, the difference in size amongst the cores of the MCF was small and unintentional in origin, so the phase changes between them were small and the scrambling caused by the coupling between the supermodes of the MCF was limited. The fibre device we present here is designed to provide improved phase and amplitude scrambling to achieve a stable near field and far field illumination output pattern during input coupling variations and under variable environmental conditions along the fibre length, and low modal noise for increased photometric stability and limited controlled FRD. It combines the coupling efficiency of a multimode fibre with modal stability similar to a single-mode fibre, with low loss across the wavelength range from 380~nm to 860~nm. Initial results of near field and far field scrambling and modal noise are presented for the specially designed MCF device and as well as those for current state of the art octagonal fibre for comparison.

\section{Design of a multicore fibre for improved scrambling and low modal noise}
\begin{figure}
\includegraphics[width=\columnwidth,scale=0.5]{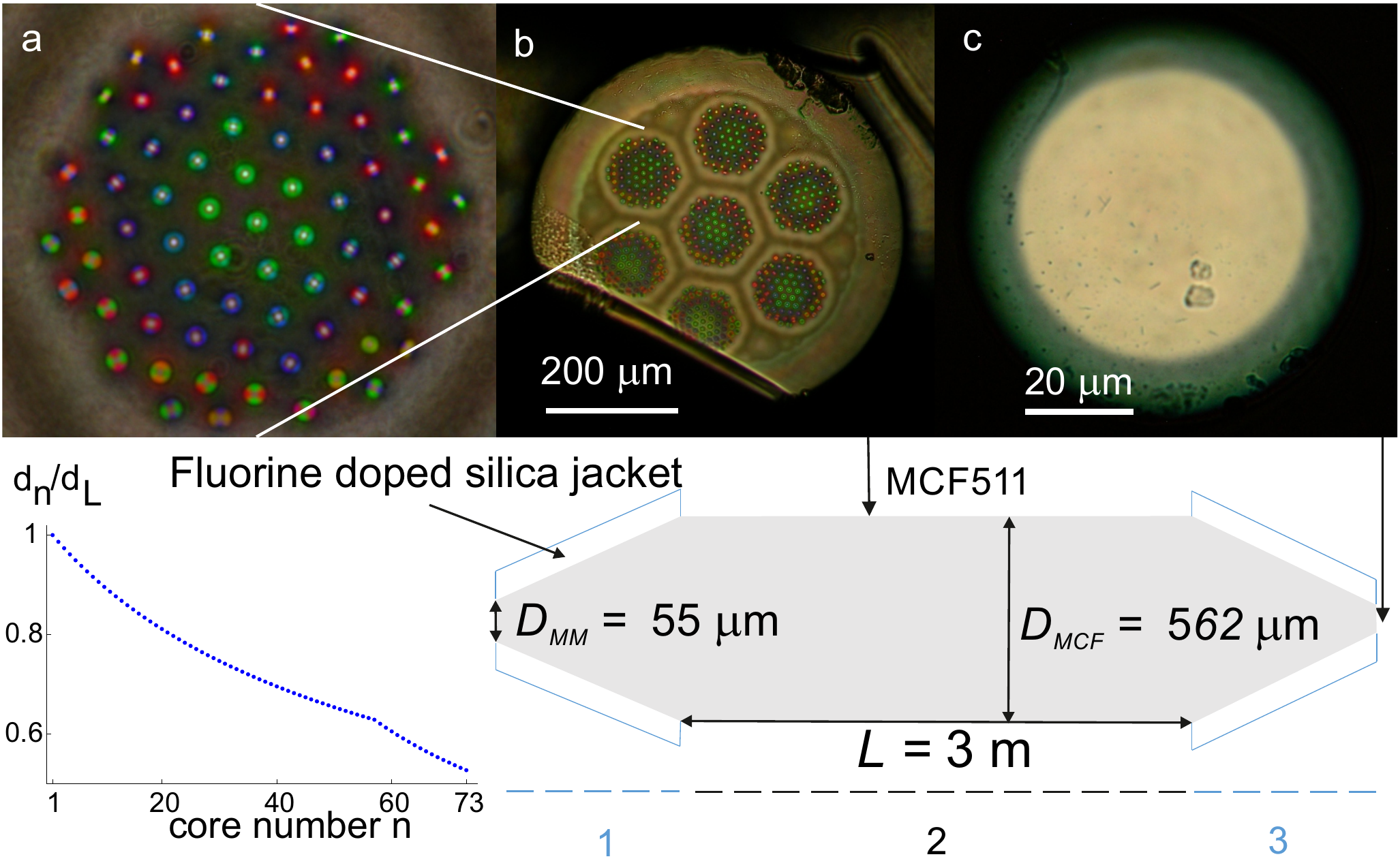}
\caption{Top: micrographs of a) detail of the MCF showing 73 different cores b) the entire 511-core MCF and c) the MM port of a photonic lantern made by tapering the MCF. Bottom left: the distribution of core diameters d$_{n}$ among the \textit{N}=73 cores, normalized to the largest core diameter d$_{L}$=3.38 $\mu$m (n=1). Bottom right: schematic of the MCF-based mode scrambler under test with three sections: 1) a photonic lantern with multimode input port that transitions adiabatically into 2) the multicore fibre of length \textit{L} = 3 m with dissimilar cores and 3) an output photonic lantern identical to the input lantern but reversed.}\label{schem_MCF_scrmblr_d_profile}
\end{figure}

\begin{figure}
\includegraphics[width=\columnwidth]{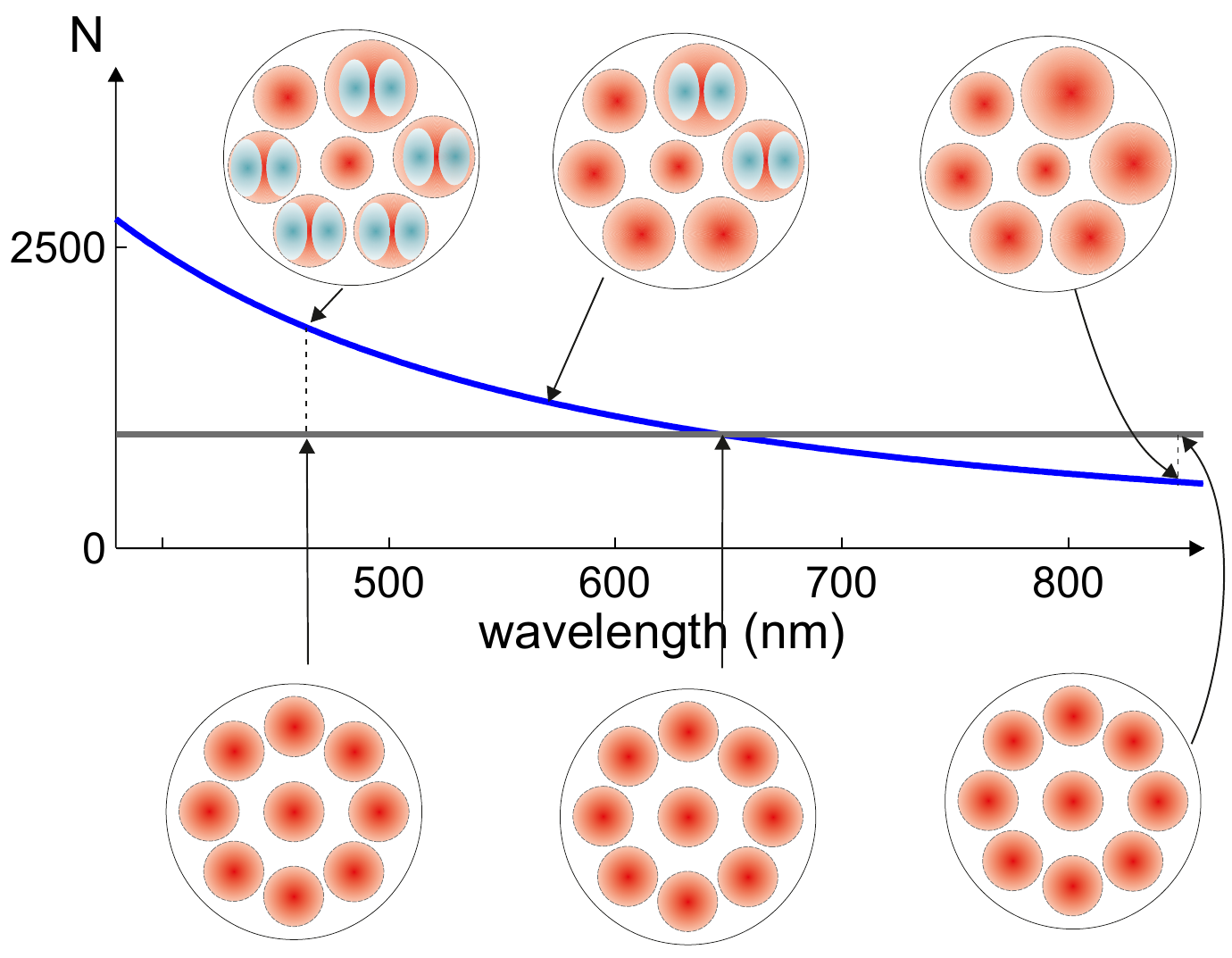}
\caption{Calculated number of guided modes N versus wavelength. The  curved line is for a representative MM core for a range of wavelengths, while the straight line is the number of modes supported by a representative MCF with identical single-mode cores. The top schematics show an MCF with dissimilar cores that are all single-mode at the longest wavelength of interest but become successively 2-moded at shorter wavelengths, such that the total number of modes matches that of the MM core across the wavelength range. For comparison, the bottom schematics show an MCF with identical single-mode cores that matches the number of modes of the MM core at only a single wavelength.}\label{Nvslambda__MCF_identical+disimilar}
\end{figure}

The broadband mode scrambler presented in this paper comprises: input and output multimode photonic lantern ports and an MCF mid-section, as illustrated in Fig.~\ref{schem_MCF_scrmblr_d_profile} (top). Unlike the MCF mentioned in the previous section, here the MCF's cores have different sizes, distributed in such a way that they support a suitably-varying number of modes across the wavelength range of interest. This has 2 advantages: 1) it enhances the phase differences between the cores and hence the scrambling of their supermodes \cite{Dionne2014} and 2) it allows broadband low-loss operation. The number of modes supported by MM fibres, such as the input and output ports of a lantern-based mode scrambler, has an inverse-square dependence on the wavelength. However, the number of cores in the MCF is fixed. Since low-loss operation of a photonic lantern relies on the number of single-mode cores matching the number of MM fibre modes, a device based on a single-mode MCF can be low loss at only a narrow band of wavelengths \cite{Birks2015}. An MCF with dissimilar cores can solve this problem if increasing numbers of the cores become successively two-moded as wavelength decreases (starting with the largest core n=1 in Fig.~\ref{schem_MCF_scrmblr_d_profile}) in such a way that the total number of MCF modes always matches the number of modes in the MM ports, as illustrated schematically in Fig.~\ref{Nvslambda__MCF_identical+disimilar}.

In principle the device does not cause FRD if the etendue of the input and output ports matches that of the instruments they interface with.

\subsection{Waveguide analysis}

The number of spatial modes (not counting the two states of polarisation separately) supported by an MM step-index core is:
\begin{equation}
N \approx\left(\frac{V_{MM}}{2}\right)^2=\left(\frac{\pi d_{MM} NA_{MM}}{2 \lambda}\right)^2 
 \label{mode_number},
\end{equation}
where $\textit{d}$$_{MM}$ is the MM core diameter and \textit{NA}$_{MM}$ is its numerical aperture at a wavelength $\lambda$. The \textit{V}$_{MM}$-number defines the number of propagating modes in the waveguide \cite{Snyder1983,Gloge1971}. 

The transition in a photonic lantern distributes the power in the modes of the MM port between the cores of the multicore region, and vice versa. Low loss is achieved by matching the number of modes in the 3 sections of the MCF mode scrambler i.e. \textit{N}$_{MM}$=\textit{N}$_{MCF}$, where \textit{N}$_{MM}$ is the number of modes in the multimode ports (Fig.~\ref{schem_MCF_scrmblr_d_profile}c) and \textit{N}$_{MCF}$ is the number of modes guided collectively by all the cores of the MCF section (Fig.~\ref{schem_MCF_scrmblr_d_profile}b) \cite{Birks2015}.

The number of modes is a convenient concept in waveguide theory, but the equivalent concept more commonly used in astronomy is etendue. The etendue of a light beam is the product of its area \textit{A} and its solid angle of divergence $\Omega$=$\pi$$\alpha$$^{2}$, where $\alpha$ is the divergence half-angle: 

\begin{equation}
A\Omega=\pi^2 \rho^2 \alpha^2
 \label{Etendue},
\end{equation}

To maximize coupling between a free-space beam and an MM optical fibre its core diameter \textit{d}$_{MM}$ and numerical aperture \textit{NA}$_{MM}$ must match those of the beam, i.e. d$_{MM}$=2$\rho$ where $\rho$ is the beam radius and \textit{NA}$_{MM}$=$\alpha$ for small angles. Using eqs. (\ref{mode_number}) and (\ref{Etendue}) one can relate the number of spatial modes \textit{N} to the etendue \textit{A}$\Omega$ as:

\begin{equation}
N=\frac{A\Omega}{\lambda^2}
 \label{N_G},
\end{equation}

\subsection{Multicore fibre design}
The MCF-based mode scrambler was designed to match a system with \textit{N}$_{L}$=480 modes (as calculated from the etendue) at $\lambda_{L}$=860~nm and an \textit{NA}$_{MM}$=0.23 at the MM ports. $\lambda_{L}$ is the longest wavelength of operation and \textit{N}$_{L}$ the corresponding number of modes of the system at  $\lambda_{L}$. If all the cores are single-moded at $\lambda_{L}$=860~nm then the fibre should have \textit{N}$_{cores}$=480 cores. However, the required operating wavelengths span the range from 380~nm to 860~nm. To match  \textit{N} with wavelength variation in Eq.~\ref{mode_number}, the wavelength $\lambda$ at which the MCF's cores collectively guide \textit{N} modes is: 

\begin{equation}
\lambda = \lambda_L\sqrt{\frac{N_{cores}}{N}} 
 \label{lambda},
\end{equation}

To match the number of modes to those of the MM core, the second-mode cutoff wavelengths of successive cores in the MCF are obtained by successively incrementing \textit{N} by 2 in Eq.~\ref{lambda}, since the second mode is in fact a degenerate pair of modes. This allows us to obtain the profile of core sizes.

There are two more subtleties to consider. Firstly, for the shortest wavelengths, the calculation is modified to take account of the fact that the biggest cores start to guide modes of even higher order. This causes the kink in the plot in Fig.~\ref{schem_MCF_scrmblr_d_profile} (bottom left). Secondly, \textit{N}$_{cores}$=480 is an awkward number of cores for our fibre fabrication process, which is based on stacking on a hexagonal lattice. Instead we adopted the more convenient value of \textit{N}$_{cores}$=511, formed from 7 clusters of 73 cores.  The core diameters in each cluster were distributed according to Eq.\ref{lambda}. Because this exceeds the number calculated from the required etendue, we correspondingly reduced $\lambda_{L}$ to 834~nm. The result is the core diameter profile shown in Fig.~\ref{schem_MCF_scrmblr_d_profile} (bottom left), such that the largest core n=1 has \textit{V}$_{co}$=2.405 (the \textit{V} value for second-mode cut-off) at $\lambda$$_{L}$=834~nm and \textit{V}=5.4 at $\lambda$=380~nm. To obtain absolute core sizes, we find the diameter \textit{d}$_{1}$ of the largest core from the requirement that it is at second-mode cutoff at $\lambda_{L}$:

\begin{equation}
d_1 = \frac{V_{co} \lambda_L}{\pi \textit{NA}_{MCF} }
 \label{d0},
\end{equation}
where the numerical aperture \textit{NA}$_{cores}$=0.19 of the MCF's cores is determined by the germanium-doped preform used to make them. Thus the core diameters should range from 1.77$\mu$m to 3.38$\mu$m.

The outer diameter of the MCF is determined by the target diameter \textit{d}$_{MM}$=52.2$\mu$m (from Eq. \ref{mode_number}) of the MM core at the narrow end of the photonic lantern. This is formed by reducing the size (i.e. tapering) of the MCF. To make the narrow end of the PL as much like a step-index MM fibre as possible, we want the residual cores of the MCF (particularly the largest one) to be an ineffective waveguide (\textit{V}$<$0.5) within the tapered-down MM core. Since the \textit{V} value of the largest core at the shortest wavelength of interest is \textit{V}=5.4, this means that the MCF needs to be scaled down at least 10 times by the tapering process, and hence that the untapered diameter of the MCF should be at least 10 times \textit{d}$_{MM}$=52.2$\mu$m. Considering practical constraints, we chose a diameter of \textit{d}$_{MCF}$=562$\mu$m. 

\section{Making the multicore fibre}

\begin{figure}
\includegraphics[width=\columnwidth]{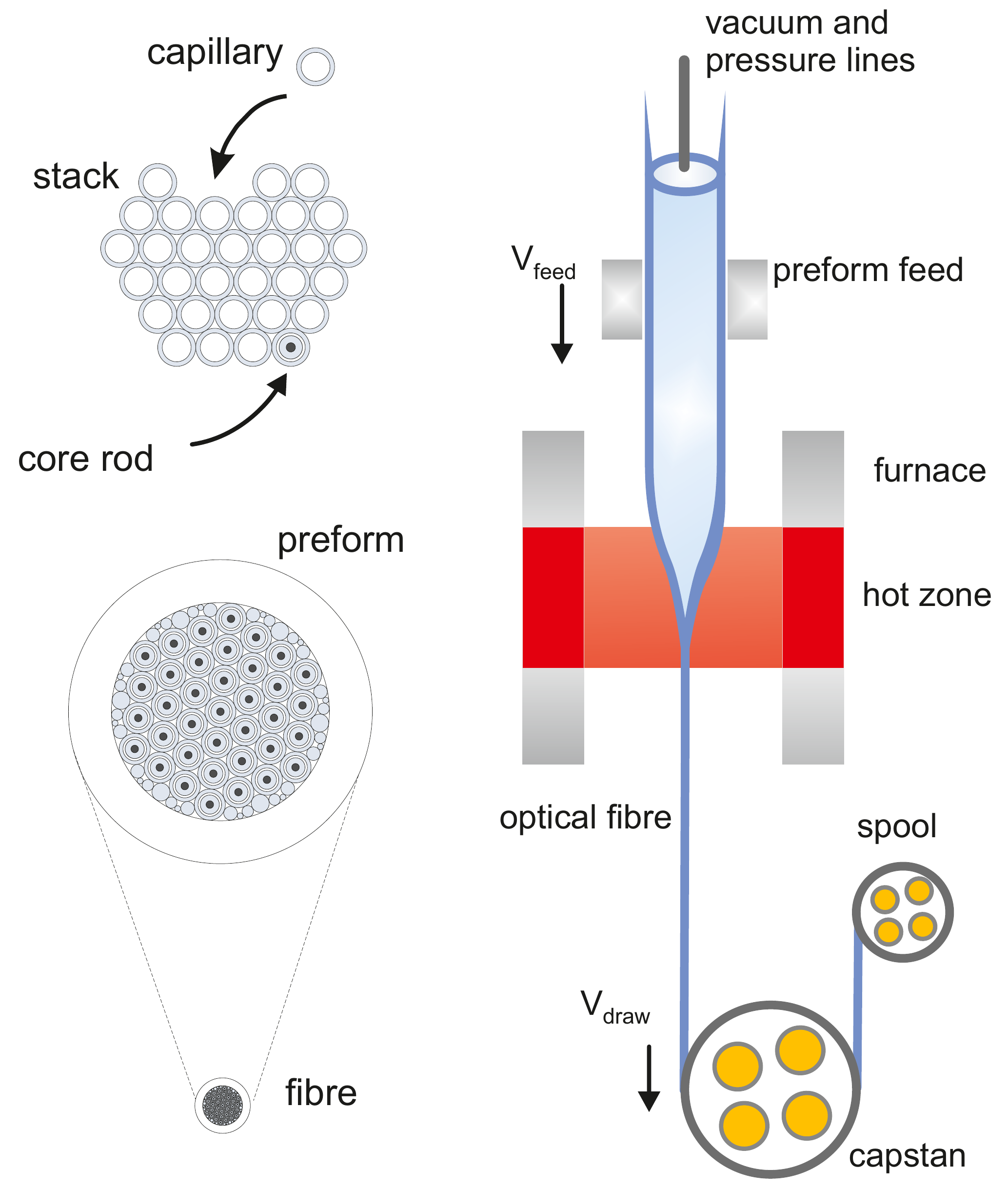}
\caption{Left: Individual capillaries and rods are stacked (top), then packed into an outer tube and drawn to fibre (bottom). In practice multiple drawing steps are used. Right: illustration of the optical fibre fabrication process and relevant parts of the fibre drawing tower.}\label{stack+tower}
\end{figure}

The 511-core fibre is fabricated in 2 stages: preform fabrication and fibre drawing.We use the stack and draw technique (commonly used to fabricate photonic crystal fibres) to build the preform, as illustrated in Fig.~\ref{stack+tower} silica glass capillaries of $\sim$1 mm diameter are drawn from a tube of outer diameter $\sim$25 mm fed into the furnace of a fibre drawing tower whilst held from the top. The bottom of the tube is kept in the furnace at $\sim$2100$^{\circ}$C. After a few minutes in the furnace the glass softens, and the lowest few centimetres of the tube will slowly sink to the bottom of the furnace under its own weight, narrowing behind it a so-called neckdown region in the furnace’s hot zone. It is here where the tube is scaled down to $\sim$1 mm in diameter. The narrowed tube is pulled from the furnace by a tractor unit and, when steady-state drawing is reached, the capillaries are cut to $\sim$1 m long pieces to form a stack. The diameter of the capillary is controlled by the mass conservation principle, changing the feed and drawspeeds of the drawing process. 73 solid germanium-doped rods, with different diameters according to the profile of Fig.~\ref{schem_MCF_scrmblr_d_profile}, are similarly drawn from a germanium-doped preform ($\sim$20 mm diameter) to make the doped rods that will form the cores of the fibre.

The capillaries are carefully stacked in a close-packed array to form a template similar to a honeycomb. The germanium-doped rods are inserted inside the capillaries, one in each. The stack is packed tightly inside a large silica tube of $\sim$25 mm diameter. So-called canes - elements with 73 cores - are made from the stack of capillaries and rods in a second stage of drawing that proceeds as before but with a vacuum line connected to the top of the stack to pump out the air between and inside the capillaries. The vacuum ensures that the canes are a completely solid. Seven of the canes are then stacked again to form a preform of 511 cores, which is then drawn down to form the final multicore fibre in a third stage of drawing. The fibre is protected with a polymer coating as it emerges from the furnace, to add mechanical protection.

For a 20 m long piece of MCF the loss measured at $\lambda$=532~nm was 0.56 dB. Higher throughput is possible for shorter lengths and/or longer wavelengths, as fibre losses typically decrease to a minimum at 1550~nm where silica glass has its minimum attenuation. 

\section{Making photonic lanterns}

The photonic lantern is a fibre device with the MCF (as fabricated) at one end and a short MM fibre waveguide at the other. The MM fibre's core is formed from the MCF after reduction in size by tapering, and its cladding is fluorine (F) doped silica, which has a lower refractive index. It is made in 2 stages using the method described in \cite{Birks2015,Birks2012}. First a short piece of F-doped silica capillary with \textit{NA}$_{MM}$=0.23 (relative to undoped silica) is threaded around the MCF. On the tapering rig, this region is heated by an oxy-butane flame (flame brushing tapering rig) or a graphite filament (Vytran tapering rig) to soften the capillary and cause it to collapse around the MCF by surface tension. Secondly this cladded structure is tapered down by heating and stretching using the same heat source, forming a symmetric biconical structure profile with a \mbox{$\sim$1 cm} long uniform waist of diameter \textit{d}$_{MM}$$\approx$73$\mu$m. The waist is cleaved to reveal the lantern's multimode port as shown in Fig.~\ref{schem_MCF_scrmblr_d_profile}c. The transition between the MCF and MM regions is $\sim$6 cm long. The diameter of the MM core is such that the largest residual cores of the MCF are not effective waveguides at the wavelengths of interest. Typical measured losses are $<$0.5 dB per lantern at 532~nm, equivalent to 89$\%$ throughput. Making more-adiabatic (longer) transitions between the MCF and the MM regions could help to achieve lower loss, ultimately increasing the overall throughput of the whole device.

\section{Scrambling and modal noise characterisation}
 	
\subsection{Experimental setup}

\begin{figure}
	\includegraphics[width=\columnwidth]{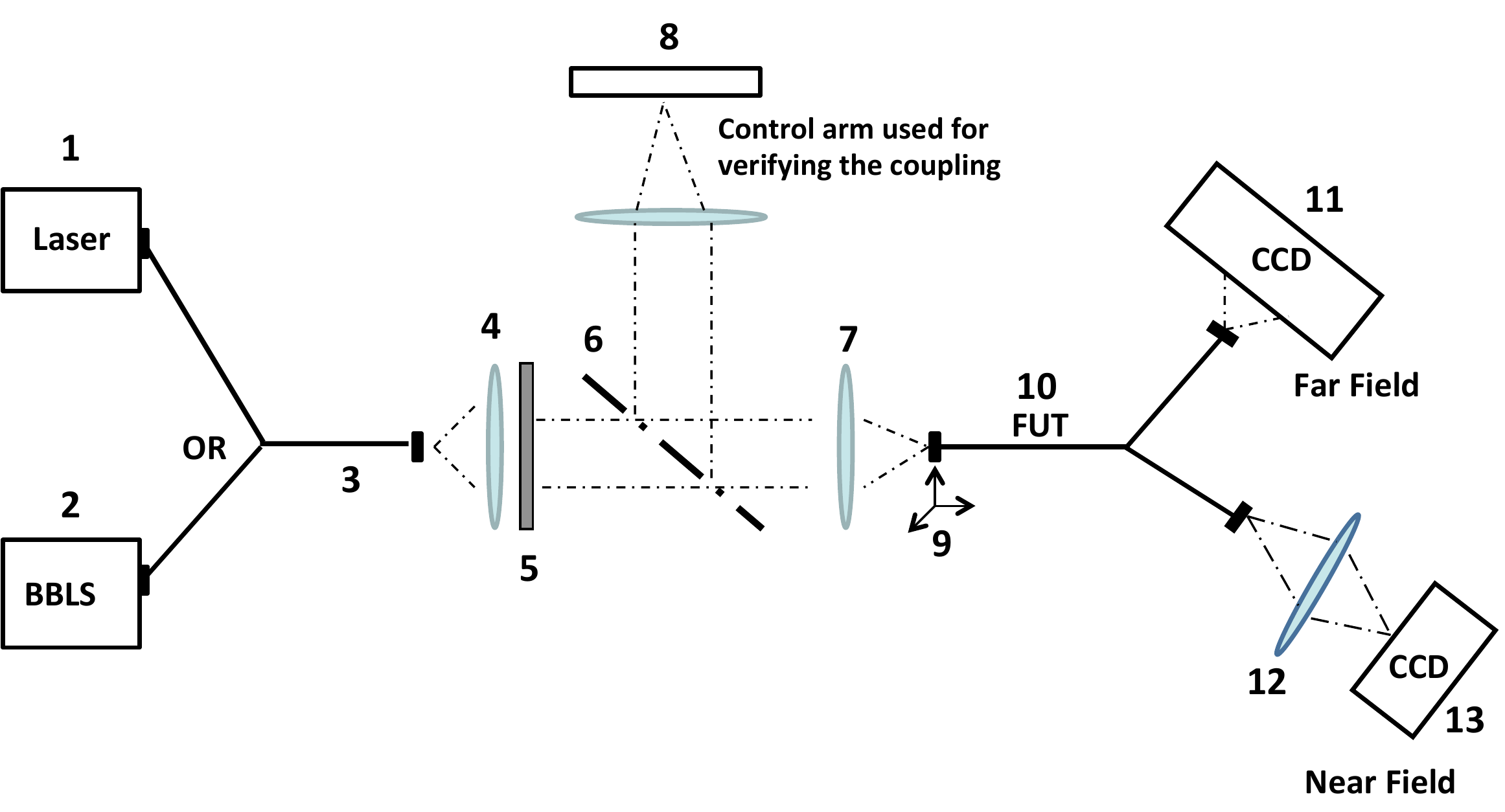}
\caption{Diagram of the experimental setup used to characterize the scrambling behavior of the multicore fibre (MCF511) and the octagonal fibre (OCT100) in the near field and far field. 1. HeNe laser 632.8~nm, 2. Broadband light source 3. Light source fibre, 4. Objective lens x10 NA 0.25, 5. Wavelength filter (central wavelengths 550~nm and 750~nm with filter FWHM of 10~nm), 6. Beam splitter, 7. Objective lens x20 NA 0.4, 8. CMOS detector, 9. 3-axis stage, 10. Fibre under test, 11. 2K x 2K CCD, 12. Objective lens x50, 13. 1K x 1K CCD.}\label{scrambling_charact_setup}
\end{figure}

\begin{figure}
	\centering
\begin{tabular}{@{}c@{}}
	\includegraphics[width=\columnwidth]{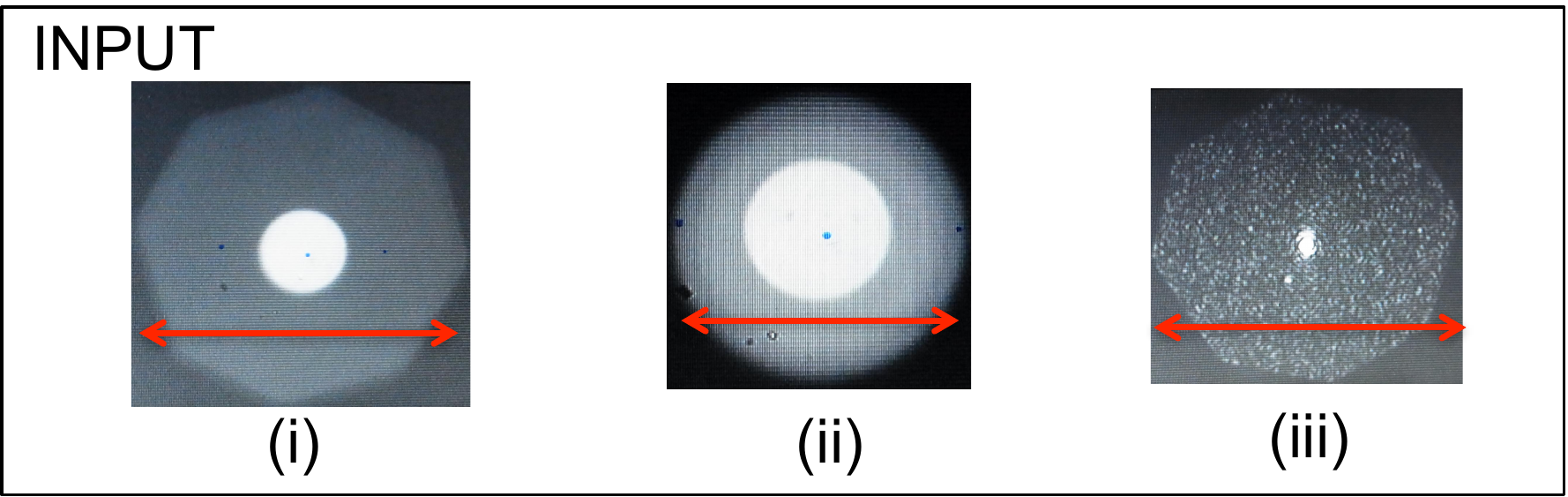}\\
	\includegraphics[width=\columnwidth]{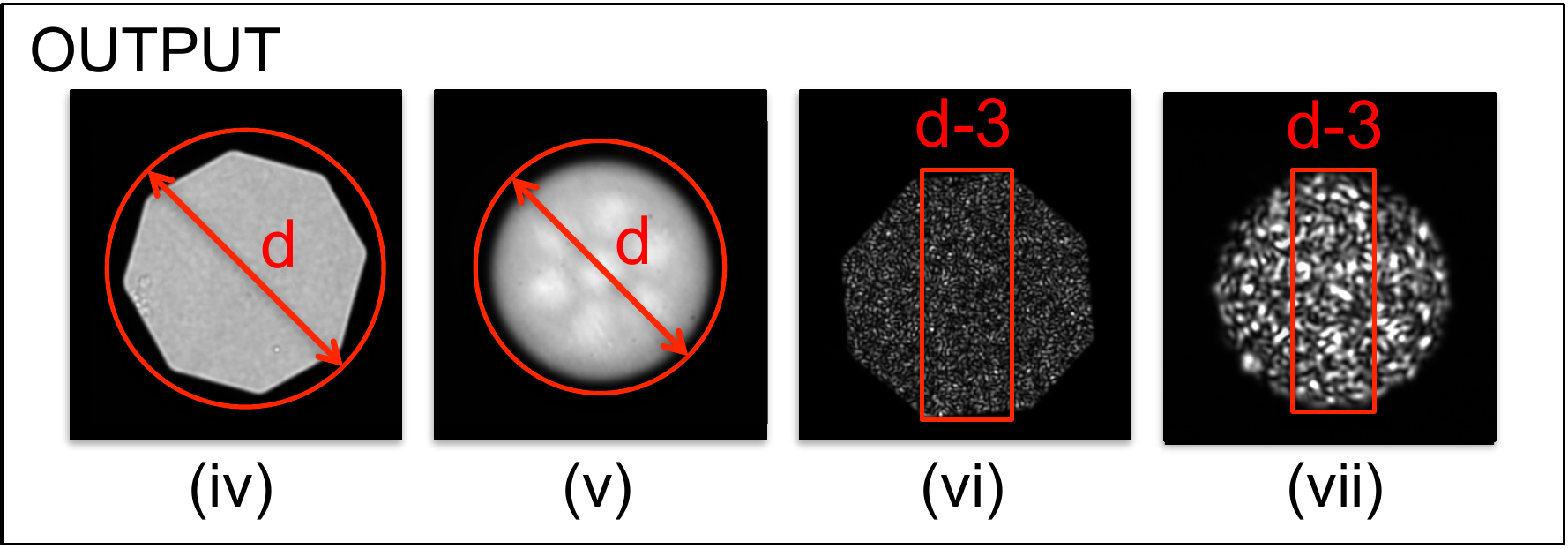}
\end{tabular}
\caption{Top: Input-coupling conditions imaged with control arm in the experimental setup. The red arrow shows the scanning direction of the input spot, the centre being 0, to the right is the positive direction and to the left the negative direction. (i) and (ii) shows incoherent light injection of an 25$\mu$m diameter spot of NA=0.4; (iii) shows coherent light injection at 632.8~nm (HeNe laser) of an 8$\mu$m diameter spot of NA=0.24. Bottom: Fibre output data analysis Regions Of Interest (ROI): full image diameter is defined as ROI-d for near field images (iv, v) and ROI-D for far field images; middle third slice simulating a slit (aperture vignetting) is defined as ROI-d-3 for near field (vi, vii) and far field.}\label{couplin_cond}
\end{figure}

The characterisation of the two fibres under test (FUT), the photonic lantern device that utilised the 511-core multicore fibre (MCF511) and the octagonal multimode fibre (OCT100), was performed using the experimental setup shown in Fig.~\ref{scrambling_charact_setup}. The MCF511 device has photonic lantern (PL) multimode (MM) input and output ports with a core diameter of \textit{d}$_{MM}$=52.2$\mu$m \textit{NA}$_{MM}$=0.23. The length of the MCF511 device is $\sim$3~m. The octagonal fibre under test (OCT100) is made by CeramOptec, product code OCT WF 100/187 P, it has a core diameter of 100$\mu$m NA 0.22 and is 3~m in length. In order to simulate variable seeing and telescope guiding errors, the input spot was moved across the face of the fibre core by the control arm, whilst monitoring the image of the input coupling conditions into the FUT (Fig.~\ref{couplin_cond}-top).

For modal noise and coherent light scrambling tests a narrowband HeNe laser (682.8~nm) was used and for incoherent light scrambling tests a tungsten halogen broadband lamp (Yokogawa AQ4305) combined with narrowband wavelengths filters (FWHM 10~nm) was used. In the absence of a laboratory high resolution spectrograph to characterise modal noise, a laser light source is employed to simulate a spectrally unresolved emission line in a high resolution spectrograph, the changing speckle pattern (modal noise) at the output end of the fibre is then monitored and analysed by using near and far field images. In order to smear out laser speckle effects resulting from the use of the coherent source the fibre was agitated (shaken) at a frequency of $\sim$20~Hz and amplitude of $\sim$1~cm. 

In order that all the modes of the of both FUT (OCT100 and MCF511) are excited, their NA is overfilled by an input spot with an NA=0.40 and diameter of $\sim$25$\mu$m which is achieved by re-imaging a 50$\mu$m core standard step-index multimode fibre illuminated with the incoherent broadband source that has been de-magnified by a factor of 2 by a microscope objective and projected onto the FUT end-face. We chose to fill the NA and under fill the fibre core area of both test fibres in order to compare scrambling performance, because under filling the fibre core area allows us to simulate the movement of a star across the fibre end face thereby simulating guiding errors and changing seeing conditions. The HeNe laser is coupled to a single mode fibre which is reimaged onto the FUT end-face to give an input spot size of approximately 8$\mu$m and NA=0.24. An input spot scan range of 30$\mu$m to 40$\mu$m, with a step size of 5$\mu$m, and 60$\mu$m to 80$\mu$m, with a step size of 10$\mu$m, for MCF511 and OCT100 respectively was achieved with the aid of an imaging system and precision 3-axis fibre feed platform. Because OCT100 has twice the core diameter of MCF511, the difference in step size was to ensure that the steps corresponded to the equivalent proportional move across the end-face of the fibre.  An APOGEE U47 1K x 1K thermoelectrically cooled CCD combined with a long working distance microscope objective (50x, NA=0.55) was used to image the near field of the FUT output end-face. Far field imaging of the FUT is achieved by directly projecting the output light onto a thermoelectrically cooled 2K x 2K APOGEE U230 CCD. In order that the intrinsic high dynamic range be retained, the images where saved in FITS file format.

\subsection{Image processing}

Using the Apogee CCD Maxim DL Pro software, the correction of the image backgrounds and bias was the first data processing step. Next, specifically tailored Python code was used to: subtract residual background, calculate image barycentre \cite{Feger2012}, and determine the integrated number of counts in selection regions of interest (ROI). In detail: each dataset consists of seven or nine frames, each with a different location of the input spot on the fibre end-face. Using the reference frame (the frame with the input spot centred position 0 on the fibre end-face) the barycentre is calculated and used as a reference point to define the centre of the circular region of interest (mask) that is used for the entire dataset. In order to simulate the aperture vignetting technique often used in high-resolution spectrographs to increase their spectral resolving power \cite{Lemke2011} ROI-d-3 digital mask (Fig. ~\ref{couplin_cond} vi and vii) was employed. This mask is applied to the entire dataset and is defined as the middle third of the circular ROI.  Barycentre shifts and integrated counts are all calculated relative to the reference frame (i.e. position 0) and the barycentre shifts are expressed as a ratio of one-thousandth of the core diameter (d/1000) \cite{Feger2012,Grupp2003}. 

\subsection{Fibre under test packaging and end-face preparation}

For secure handling, the input and output sections of both OCT100 and MCF511 were protected in silica capillary tubing 10~cm and 15~cm long respectively. Specially formulated low shrinkage and low strain EpoTek 301-2 epoxy adhesive was used to secure the tubing and ensure minimal potential inducted strain on the fibres. High optical quality ends with minimal subsurface damage and low surface roughness \cite{Haynes2011} were carefully prepared using a wet lapping, polishing and inspection technique.

\section{Near field results}
\subsection{Incoherent light} \label{nf_results}

\begin{figure}
	\includegraphics[width=\columnwidth]{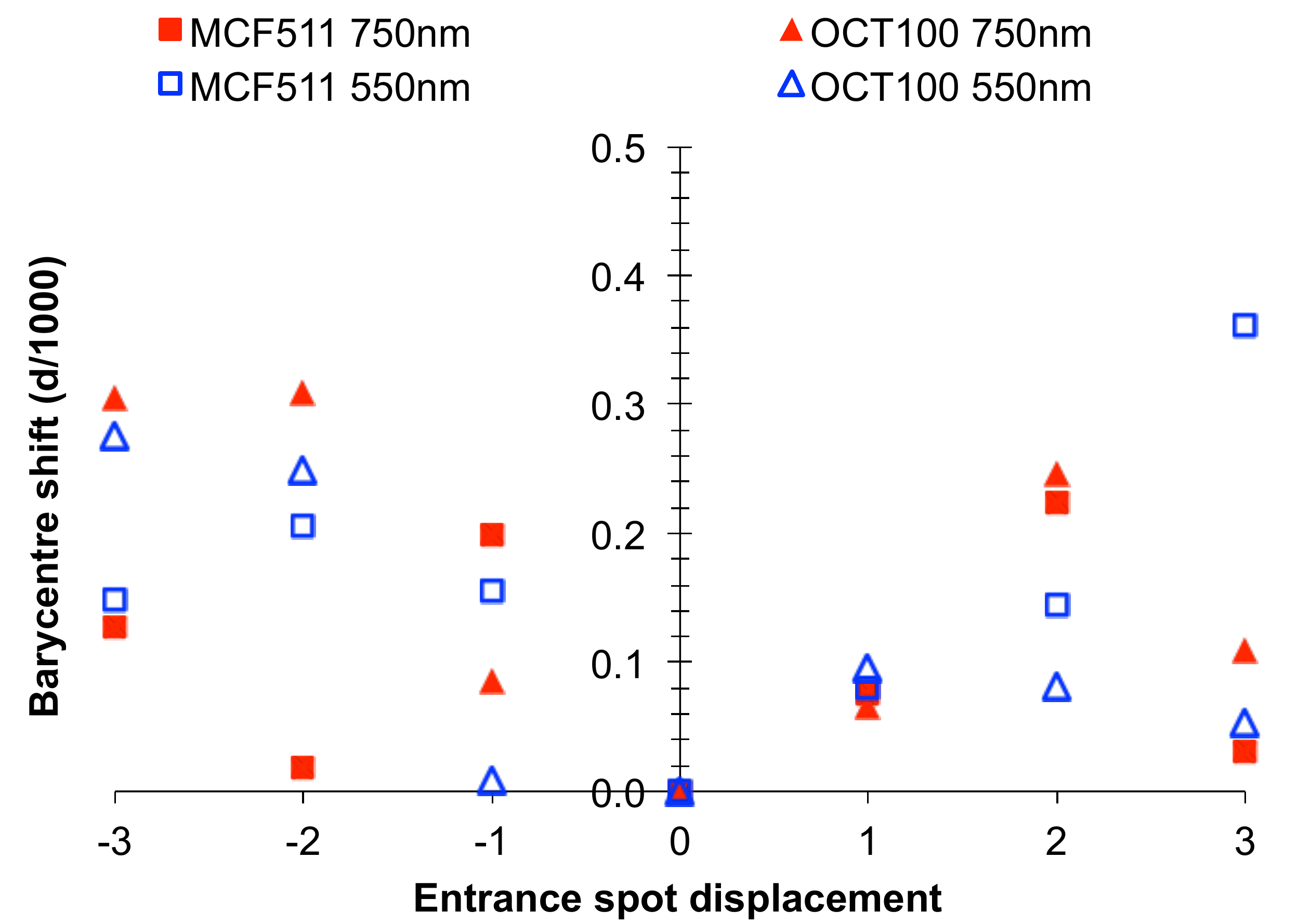}
\caption{Near field geometrical barycentre shifts for MCF511 and OCT100 FUT. Incoherent light input feed spot of 25~$\mu$m diameter at central wavelengths of 550~nm (FWHM~10nm) and 750~nm (FWHM 10~nm). The entrance spot displacements (horizontal axis) are in units of 5 $\mu$m (MCF511) and 10~$\mu$m (OCT100). Fibres are static, i.e. no shaking.}\label{nf_shift}
\end{figure}

\begin{figure}
	\includegraphics[width=\columnwidth]{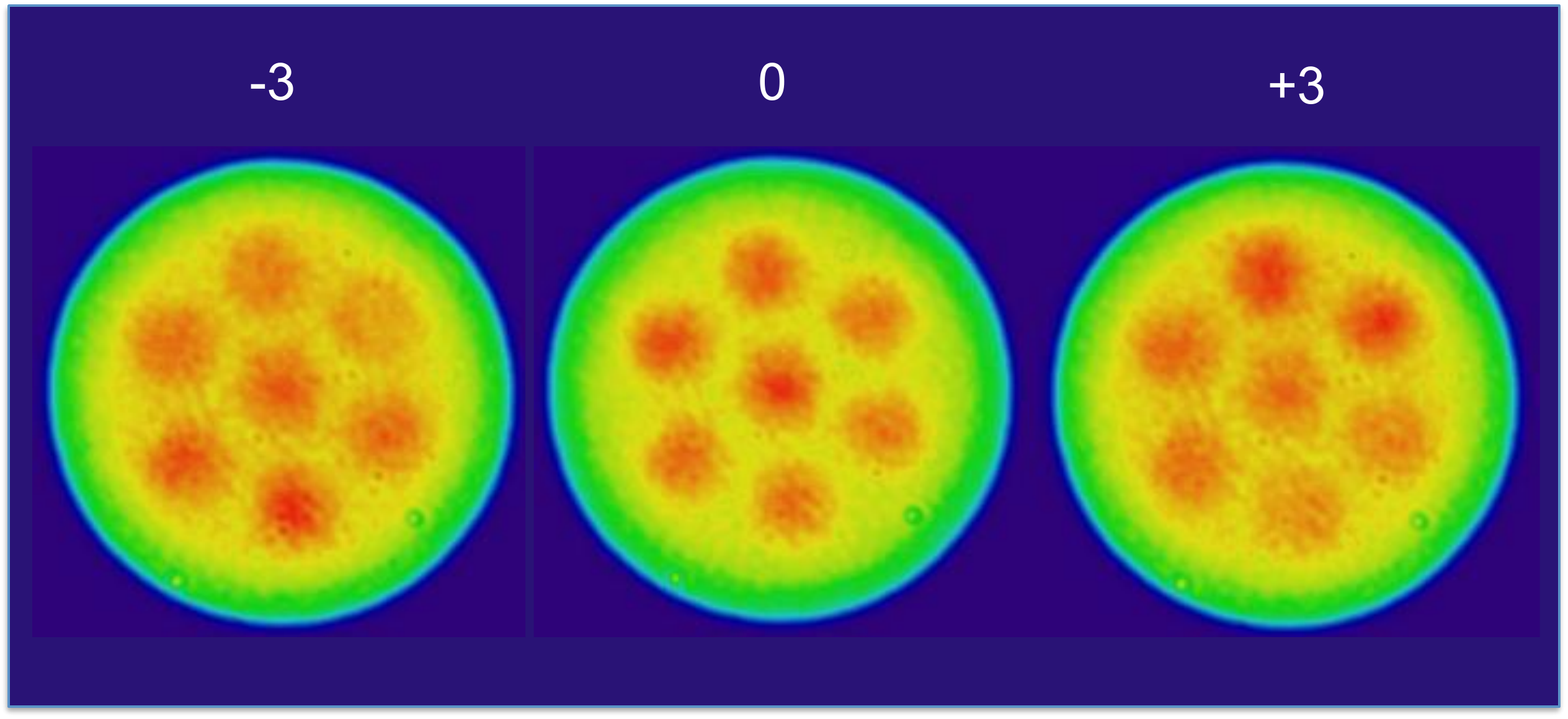}
\caption{Near field images of MCF511 PL MM output port at 550~nm for entrance spot position 0 (centre), 
and $\pm$3 ($\pm$15$\mu$m from centre position). The 7 groups of 73 cores appear to be acting as weak waveguides.}\label{nf_550nm}
\end{figure}

It has been shown \cite{Baudrand2001,Feger2012,Grupp2003,Lemke2011,Rawson1980},  that the scrambling performance of a multimode fibre improves with increasing number of propagating modes.  For our tests we fill the NA of both fibres thereby exciting all modes. Using Eq. \ref{mode_number} We calculate that at 550~nm OCT100 has $\sim$3947 modes excited, MCF511 $\sim$1175 modes excited and at 750~nm OCT100 $\sim$2123 modes excited and MCF511 $\sim$632 modes excited. Based on the mode numbers it is therefore expected that the OCT100 fibre has better scrambling performance than the MCF511 fibre, and that the OCT100 should perform better at 500~nm than 750~nm.  Fig.~\ref{nf_shift} shows the geometrical barycentre shifts measured for OCT100 and MCF511 when the input feed spot is moved across the input-face of the fibre core. It can be seen that for OCT100 the barycentre shifts are smallest at 550~nm which agrees with previous findings, however MCF511 which has ~3.3 times less modes than OCT100 has similar performance to OCT100 at both wavelengths (except for +3 position at 550~nm) suggesting that the MCF511 device is a more efficient scrambler for a reduced number of modes. It is likely that the largest barycentre shift for MCF511 at 550~nm (+3 position) is caused by remaining residual core structure in the MM end of the photonic lantern (PL). Fig.~\ref{nf_550nm} shows the near field images of MCF511 at 550~nm. The 7 clusters of 73 cores when tapered down to form the PL MM core appear to be acting as weak waveguides. 

\subsection{Coherent light}\label{Coherent_light}
  
\begin{figure}
\centering
\begin{tabular}{@{}c@{}}
	\includegraphics[width=\columnwidth]{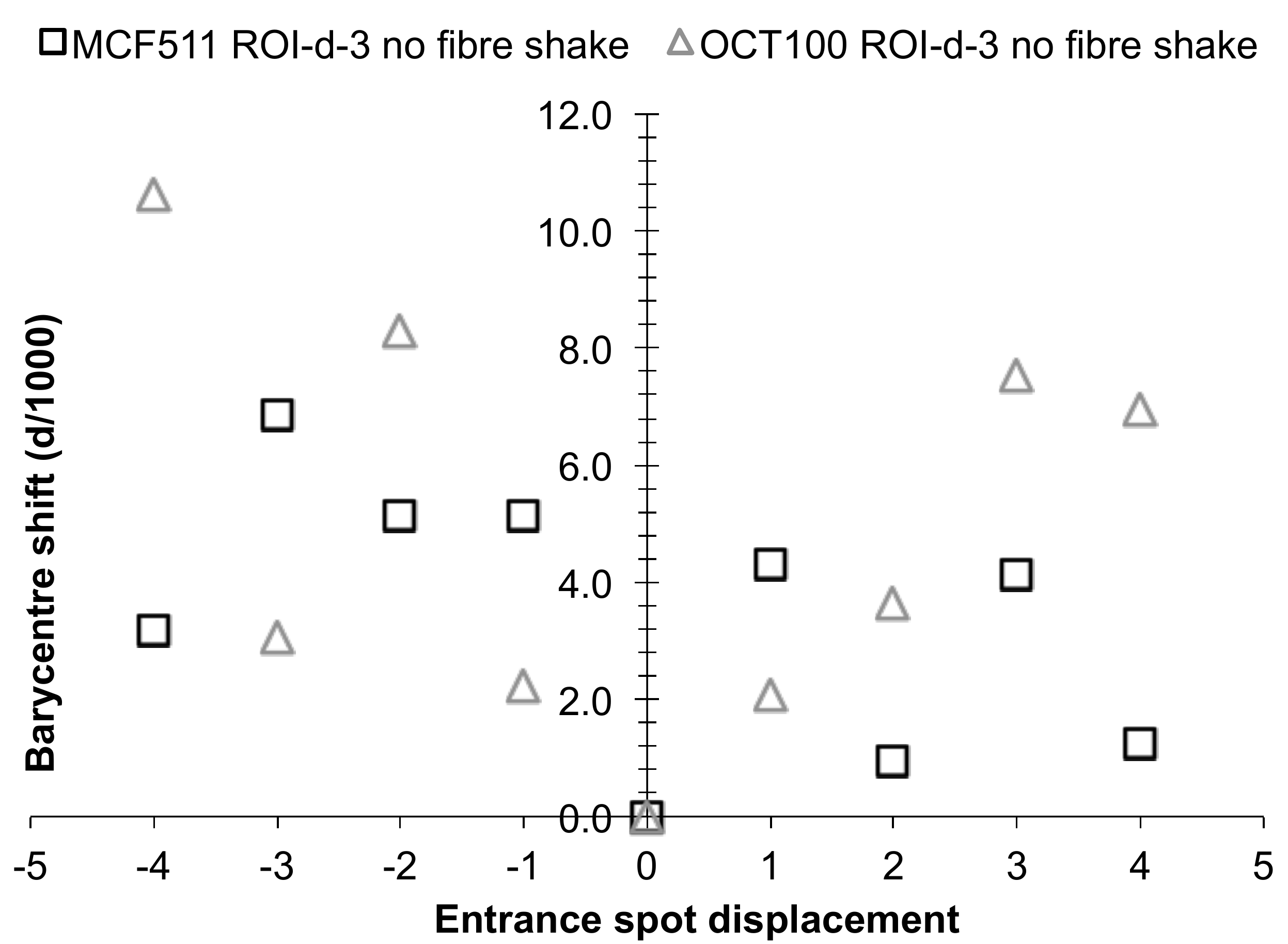}\\
	\includegraphics[width=\columnwidth]{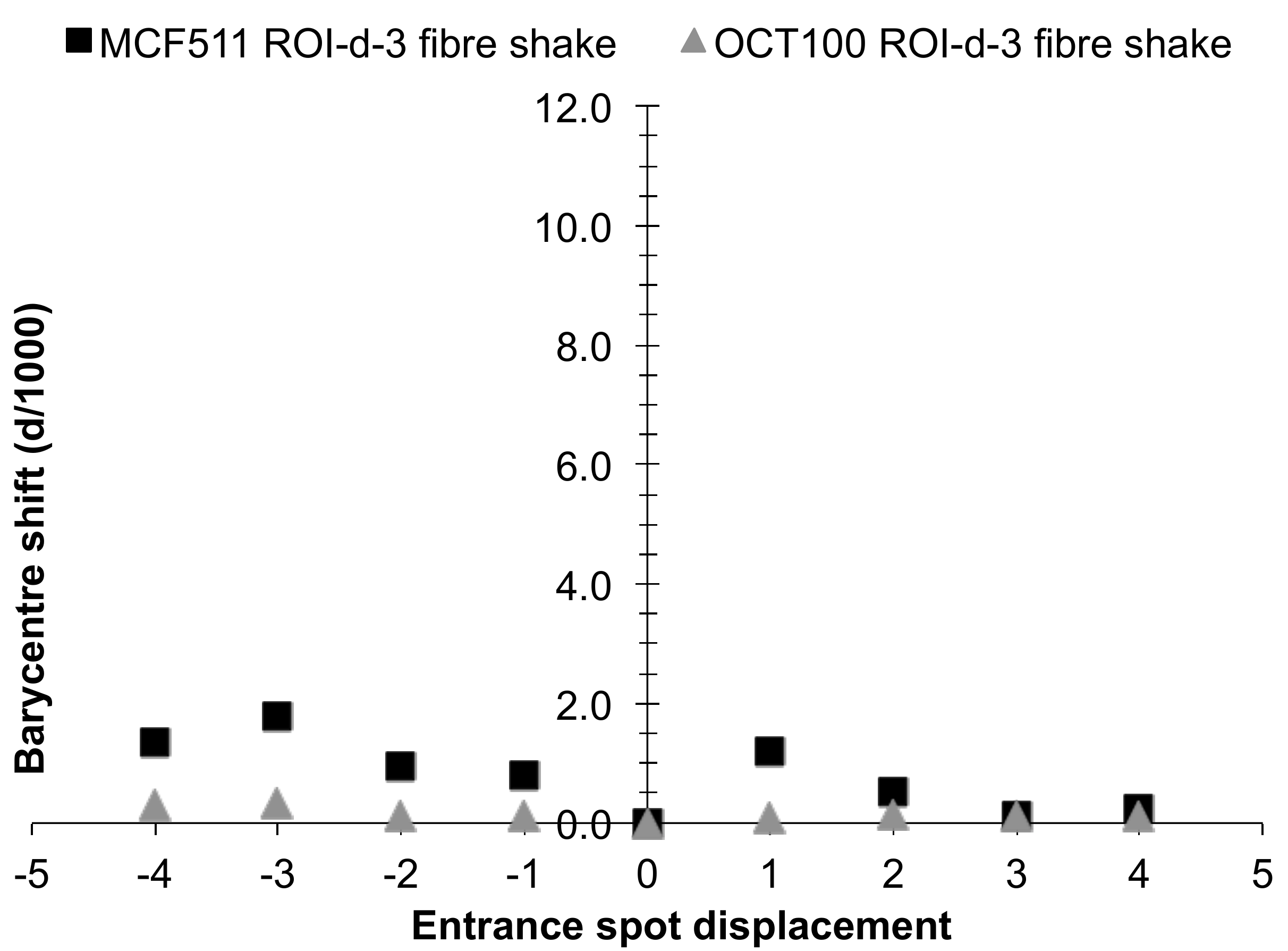}
\end{tabular}
\caption{Near field geometrical barycentre shifts resulting from scanning an 8~$\mu$m diameter input spot of coherent light at 632.8~nm across the 	FUT cores with aperture vignetting applied (ROI-d-3). Top: no fibre shake. Bottom: fibre shake.}\label{nf_geom_barycenter}
\end{figure}

An $\sim$8$\mu$m spot from a HeNe laser (632.8~nm) is scanned across the core of the FUT in order to perform the coherent light based extreme phase and amplitude scrambling tests. Measurements were taken with the fibre static and with it being shaken, the later was to reduce the impact of instabilities introduced in the speckle pattern from the fibre and improve the image scrambling performance.  Aperture vignetting \cite{Lemke2011} is applied by using the mid third section (ROI-d-3) of the near field image of the FUT core, and the normalised integrated counts and barycentre shifts are calculated and used as a measure of modal noise.

\begin{figure}
\centering
\begin{tabular}{@{}c@{}}
	\includegraphics[width=\columnwidth]{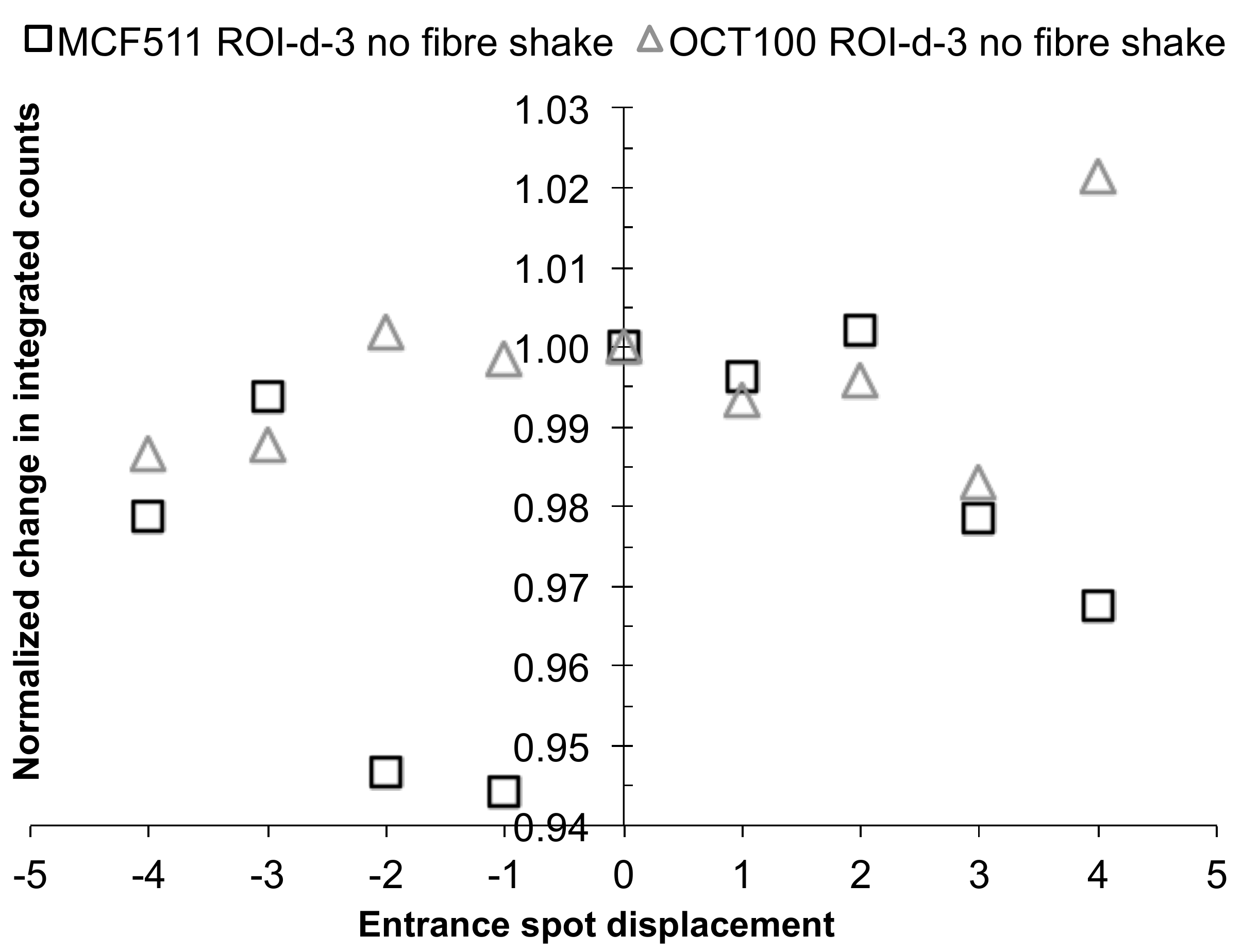}\\
	\includegraphics[width=\columnwidth]{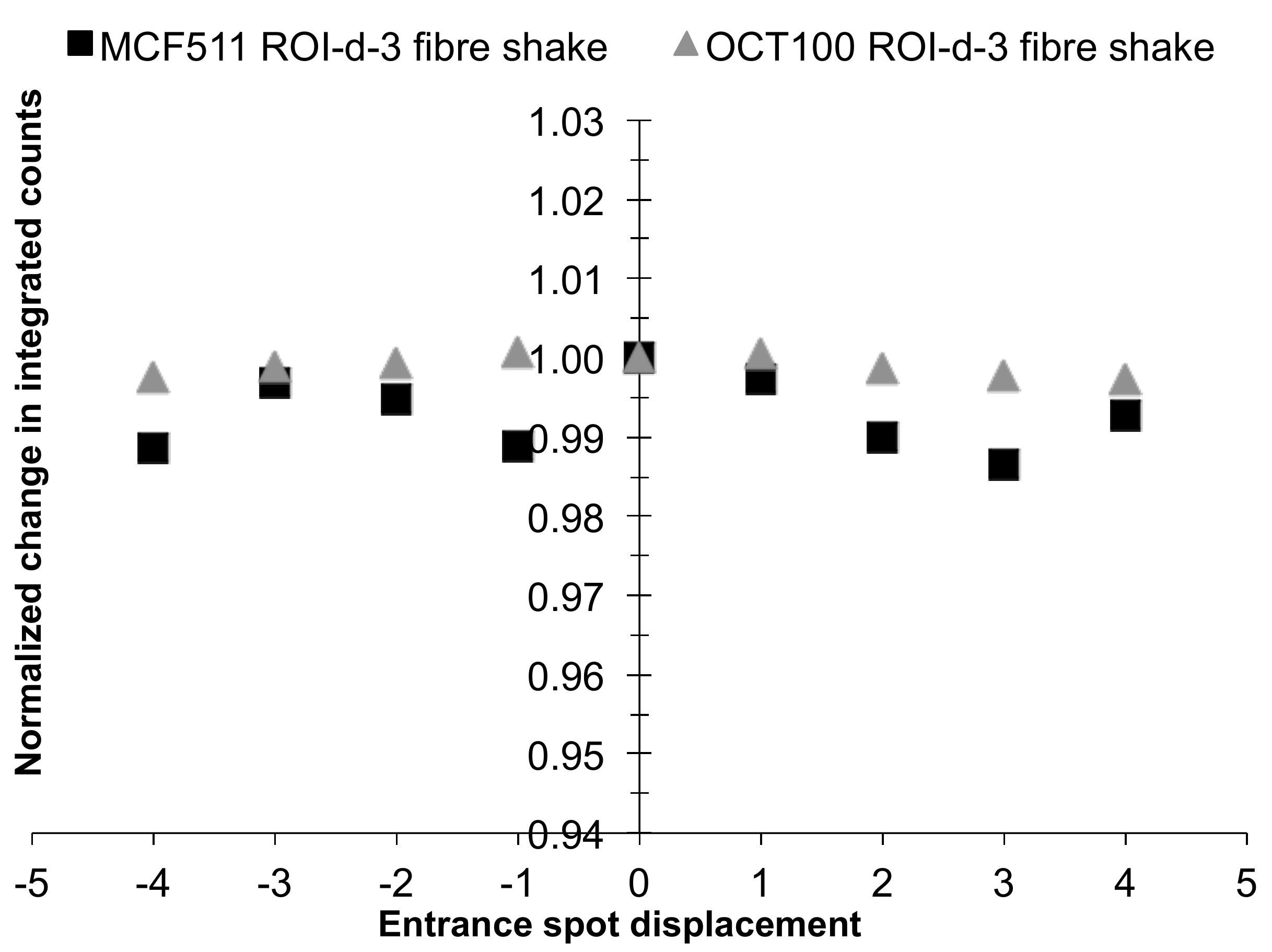}
\end{tabular}
\caption{Near field change in integrated counts (modal noise) resulting from scanning an 8~$\mu$m diameter input spot of coherent light at 632.8~nm across the FUT cores with aperture vignetting applied (ROI-d-3). Top: no fibre shake. Bottom: fibre shake.}\label{modal_noise}
\end{figure}

\begin{figure}
\centering
\begin{tabular}{@{}c@{}}
	\includegraphics[width=\columnwidth]{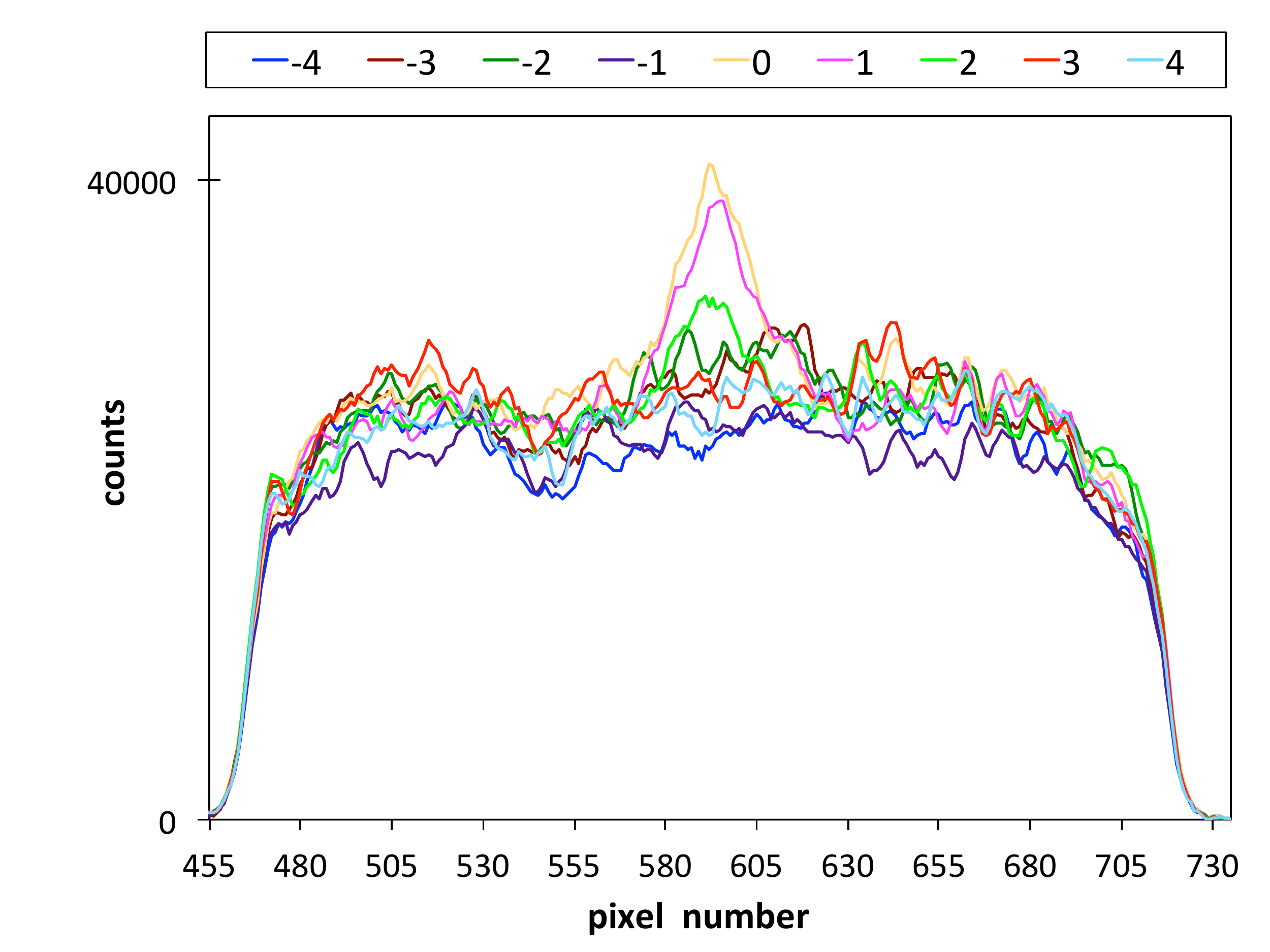}\\
	\includegraphics[width=\columnwidth]{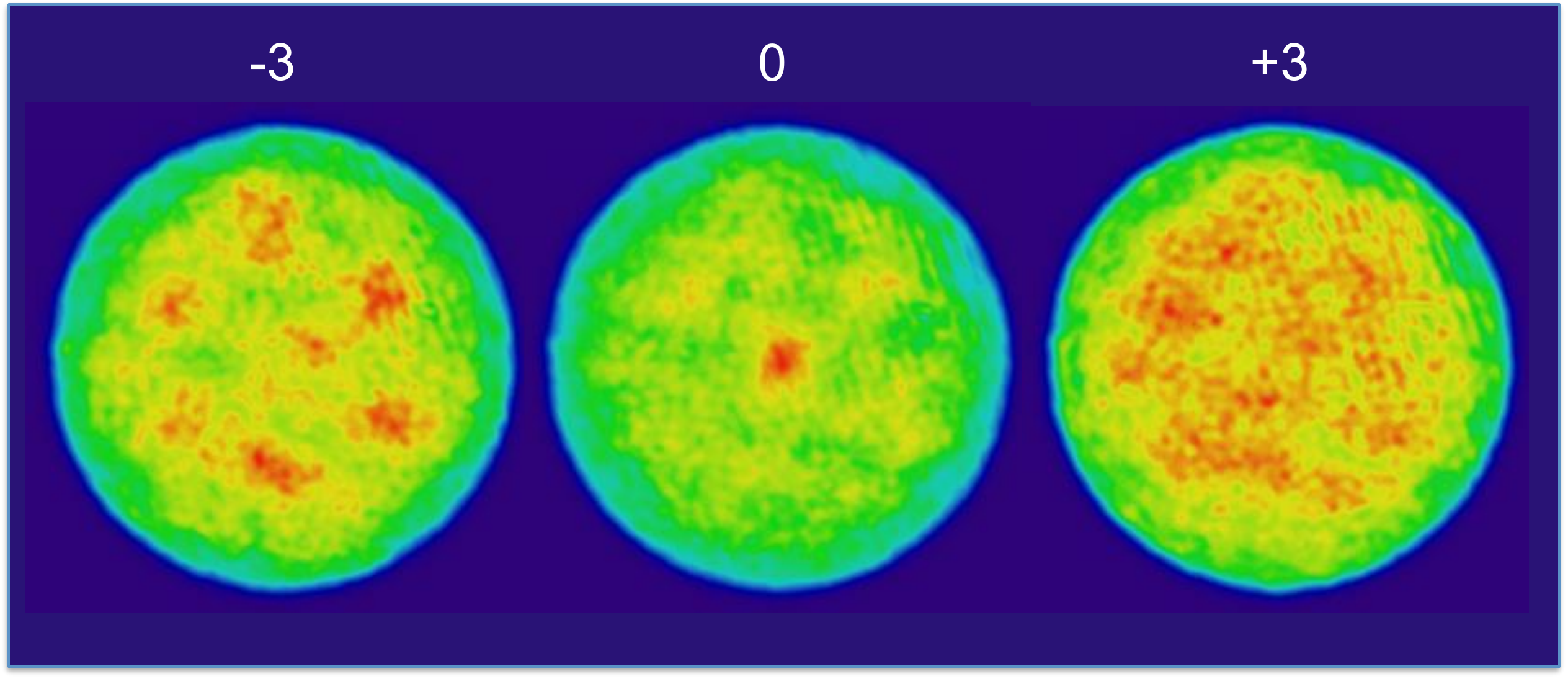}
\end{tabular}
\caption{Top: MCF511 near field 1-D slices in the x-direction (for laser input coupling and fibre shake) at each measured barycenter for the 9 input coupling positions. Bottom: Near field images of MCF511 PL MM output port, for laser input coupling and fibre shake.}\label{nf_spectr_barycenter}
\end{figure}

Using Eq. ~\ref{mode_number} we calculate that at 632.8~nm OCT100 has $\sim$2982 modes excited and MCF511 $\sim$888 modes excited, again it is expected that OCT100 should have improved scrambling and lower modal noise due to a higher number of excited modes. Fig.~\ref{nf_geom_barycenter} shows the geometrical barycentre shifts measured when aperture vignetting is applied (mid third slice of fibre core) and Fig.~\ref{modal_noise} shows the measured change in integrated counts for situations where the fibre is static (no fibre shake) and dynamic (fibre shake). Again, despite the reduced number of modes the MCF511 performance is comparable to the OCT100 in terms of scrambling and modal noise, however as mentioned previously in section ~\ref{nf_results} we suspect that the performance of MCF511 is limited by the residual core structure acting as weak waveguides. Fig.~\ref{nf_spectr_barycenter} shows near field images of MCF511 for laser input coupling and fibre shake as well as a plot showing the 1-D slice at each measured barycentre.
 
The effects of the weak waveguides could be removed by splicing a short section of circular MMF onto the end of the PL. We are currently working on this MCF511-PL modification and will retest the modified MCF511-PL in the future. At the extreme spot locations (i.e. $\pm{4}$) in the 1-D slice shown in Fig.~\ref{nf_spectr_barycenter}, the drop in counts likely results from light coupling into cladding modes.

\section{Far field results} 

The far field patterns generated by the MCF511 and the OCT100 fibre devices, when either a coherent or incoherent light spot was scanned across the fibre core, were recorded by directly projecting the output onto a thermoelectrically cooled 2k x 2k CCD. To measure the modal noise in the far field we again use the coherent light source (HeNe laser) and the aperture vignetting method detailed in section ~\ref{Coherent_light}. 

\subsection{Incoherent light}

\begin{figure}
	\includegraphics[width=\columnwidth]{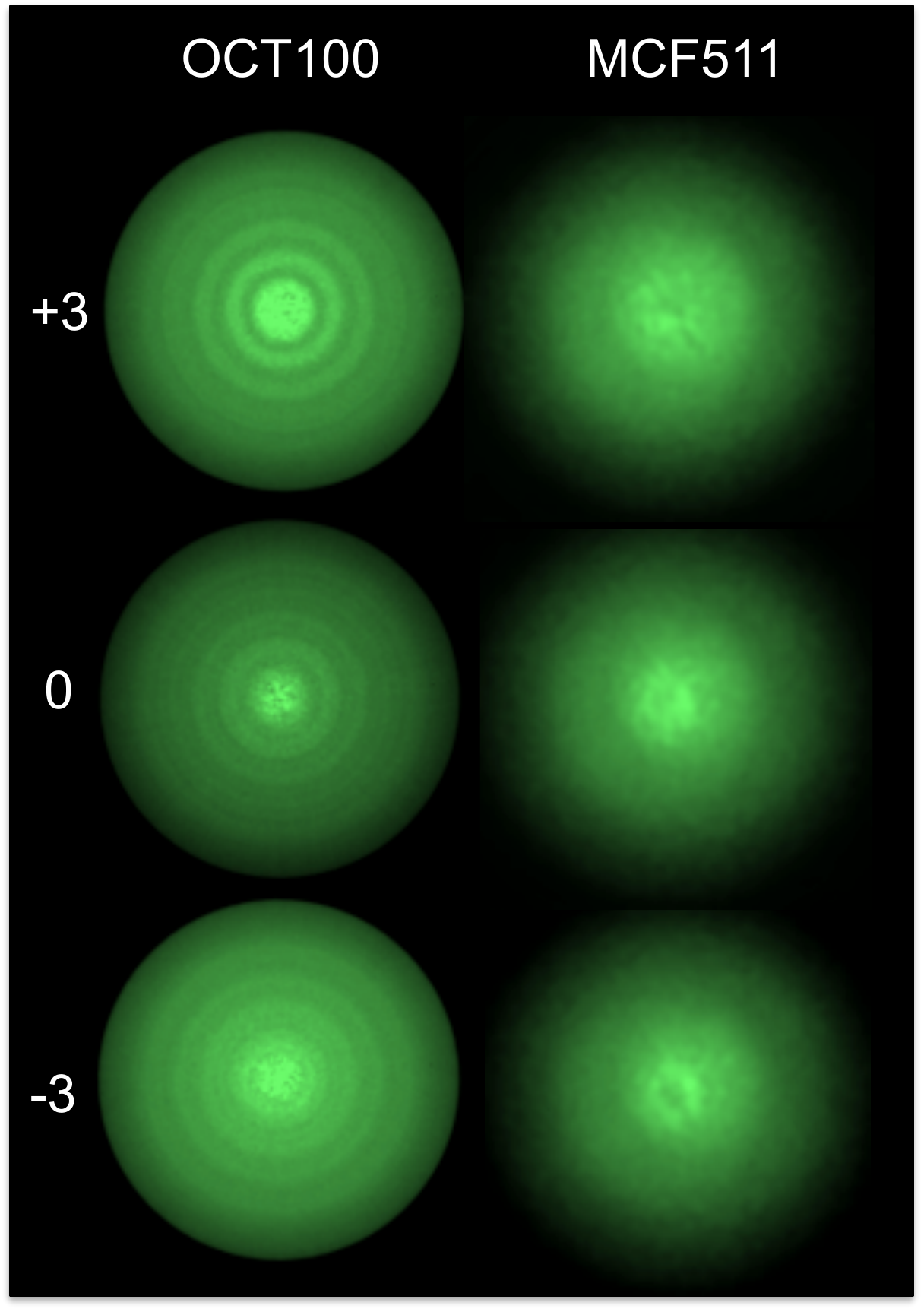}
\caption{Far field patterns from OCT100 and MCF511 FUTs resulting from the following input coupling conditions: incoherent light $\lambda$=550~nm (FWHM 10~nm), input spot positions 0, and $\pm$3, input spot diameter $\sim$25$\mu$m.}\label{ff_displacements_550nm}
\end{figure}

\begin{figure}
	\includegraphics[width=\columnwidth]{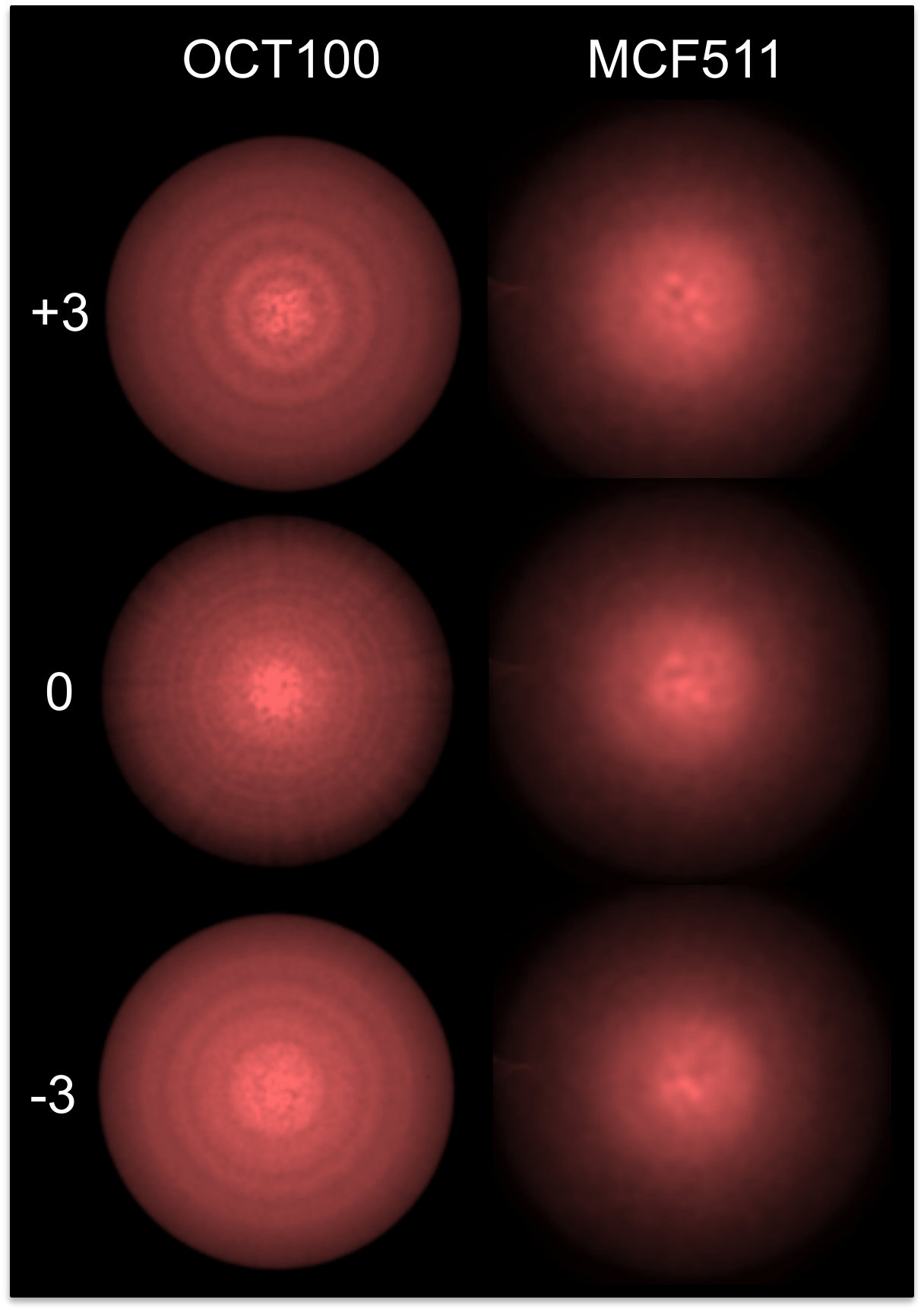}
\caption{Far field patterns from OCT100 and MCF511 FUTs resulting from the following input coupling conditions: incoherent light $\lambda$=750~nm (FWHM 10~nm), input spot positions 0, and $\pm$3, input spot diameter $\sim$25$\mu$m.}\label{ff_displacements_750nm}
\end{figure}

The far-field patterns for MCF511 and OCT100 shown in Figs. ~\ref{ff_displacements_550nm} and ~\ref{ff_displacements_750nm} correspond to the input spot locations of 0, +3 and -3 which are the centre of the core, +30$\mu$m for OCT100 or +15$\mu$m for MCF511 and -30$\mu$m for OCT100 or -15$\mu$m for MCF511 respectively. The highly distinctive Airy pattern present in the far-field distribution of the OCT100 fibre is the characteristic Fraunhofer diffraction pattern you get from a top-hat illumination function in the fibre near field \cite{Zhu2010}. Because the far field typically illuminates the spectrograph grating and much of its optics, to reduce scattered light and minimise aberrations a stable Gaussian distribution in the far field is highly desirable \cite{Ellis2014}.

\subsection{Coherent light}

\begin{figure}
	\includegraphics[width=\columnwidth]{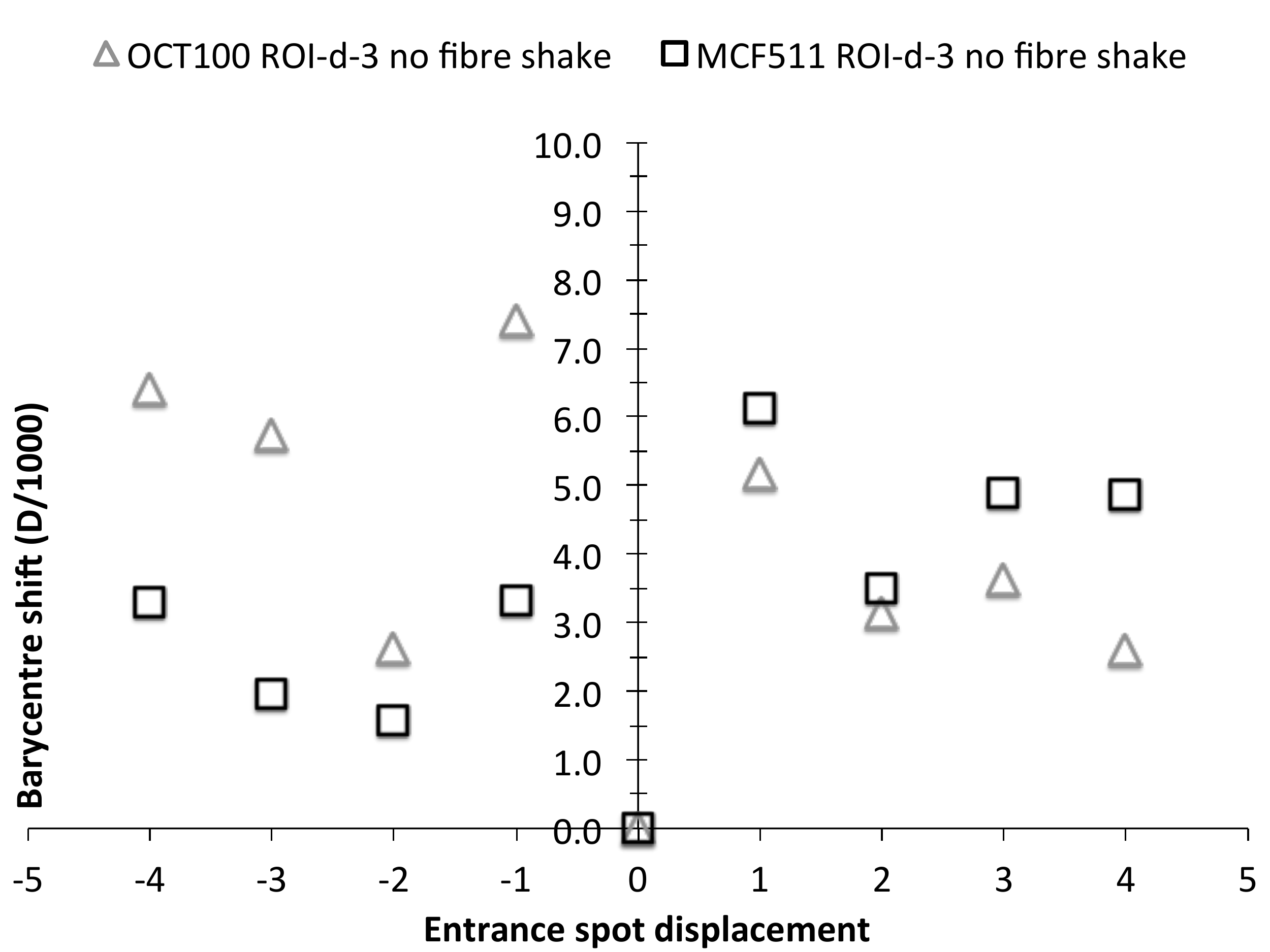}
\caption{Far field geometrical barycentre shifts resulting from scanning an 8 $\mu$m diameter input spot of coherent light at 632.8~nm across the FUT cores with aperture vignetting applied (ROI-d-3) and no fibre shake.}\label{ff_geom_barycenter_noshake_5um}
\end{figure}

\begin{figure}
\centering
\begin{tabular}{@{}c@{}}
	\includegraphics[width=\columnwidth]{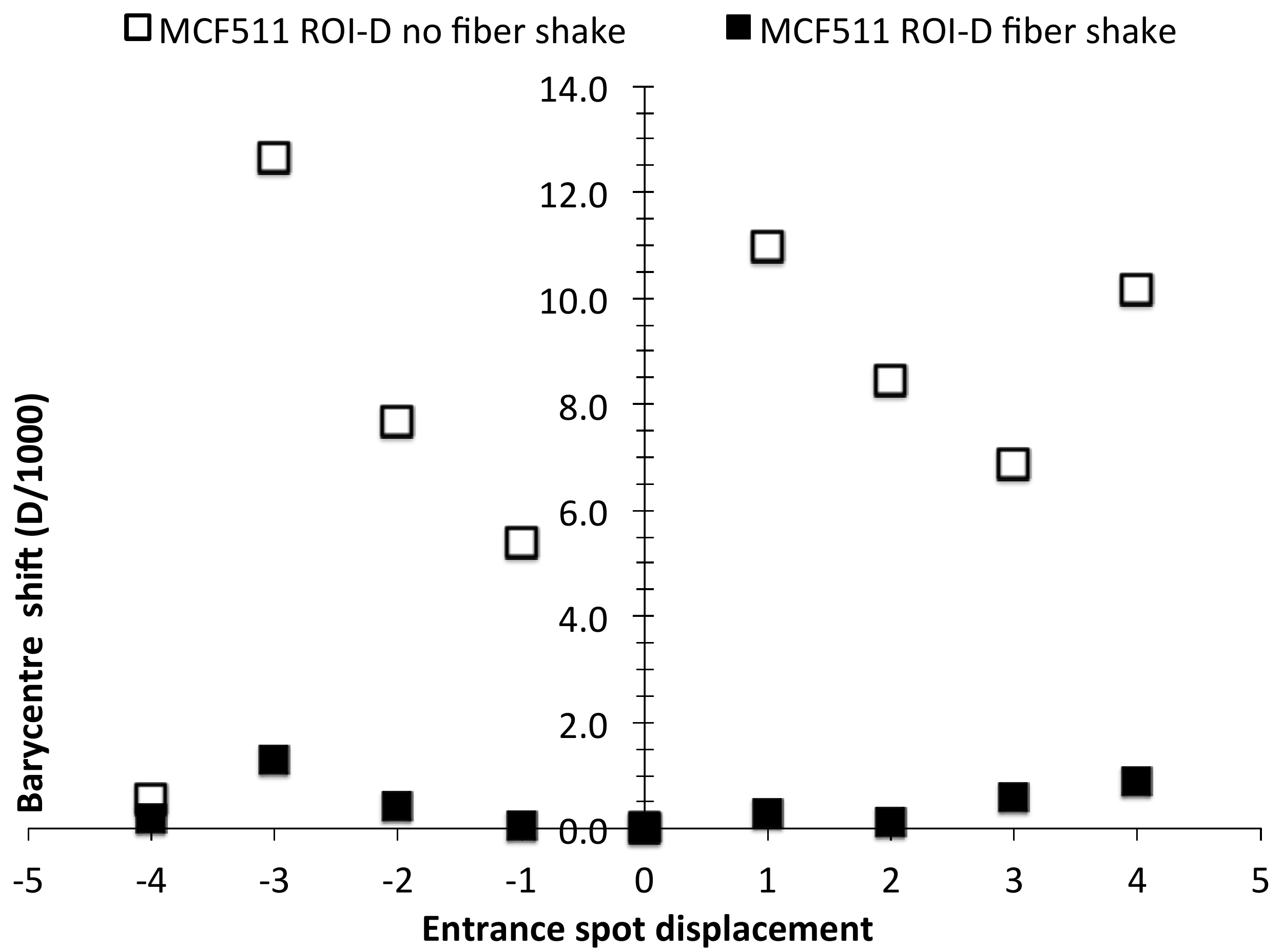}\\
	\includegraphics[width=\columnwidth]{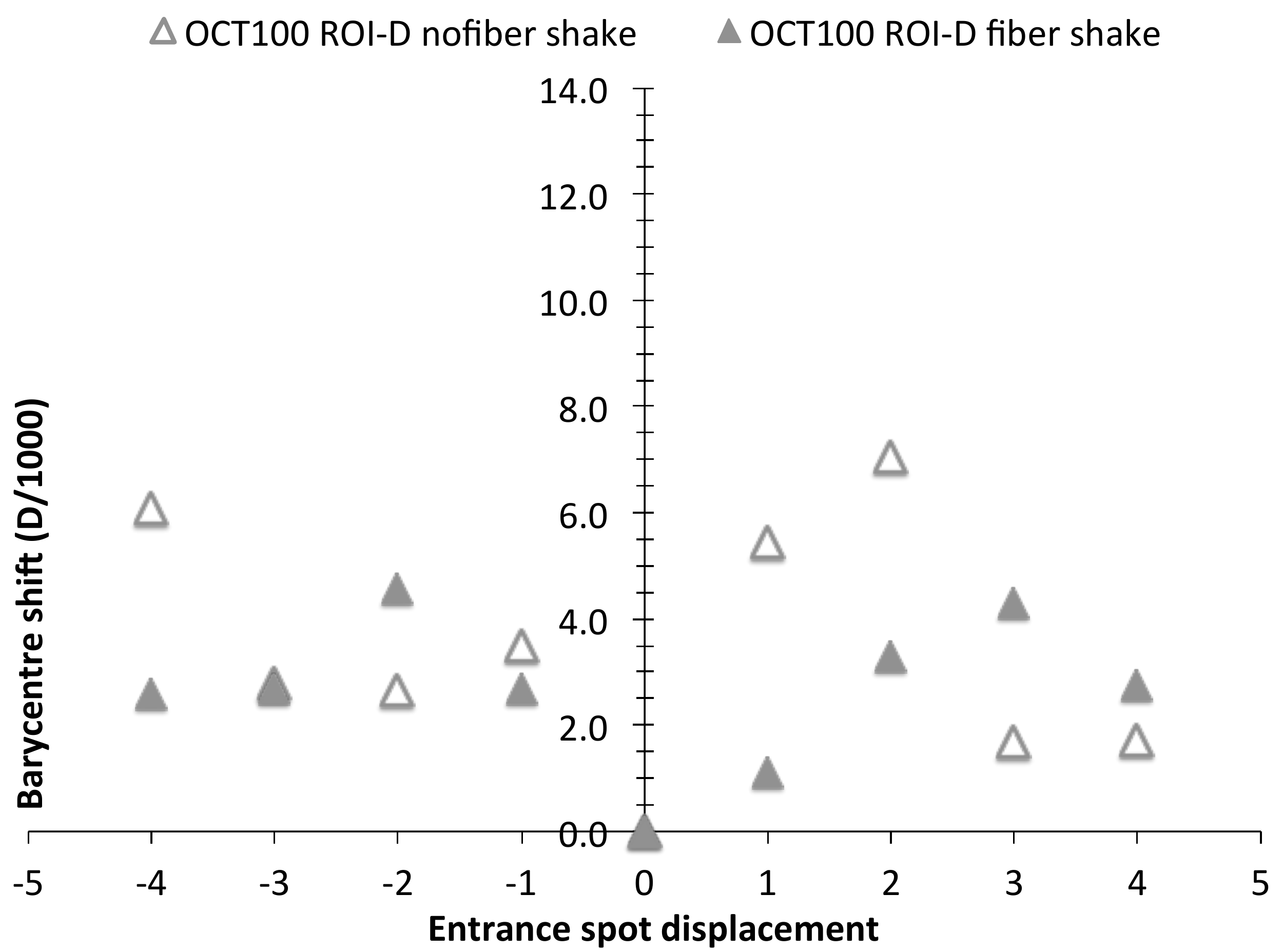}
\end{tabular}
\caption{Far field geometrical barycentre shifts resulting from scanning an 8 $\mu$m diameter input spot of coherent light at 632.8~nm across the FUT cores with no aperture vignetting applied (ROI-D). Comparison of FUT performance for fibre shake and no fibre shake. Top: MCF511. Bottom: OCT100.}\label{ff_geom_barycenter_noshake_8um}
\end{figure}

In Fig.~\ref{ff_geom_barycenter_noshake_5um}, the far-field scrambling efficiency is slightly better for MCF511 than for OCT100, with aperture vignetting and no fibre shake applied. In (Fig.~\ref{ff_geom_barycenter_noshake_8um}), without aperture vignetting, the OCT100 fibre far-field scrambling efficiency demonstrates no significant improvement as a result of shaking, whereas the MCF511 fibre significantly improves. 

\begin{figure}
\centering
\begin{tabular}{@{}c@{}}
	\includegraphics[width=\columnwidth]{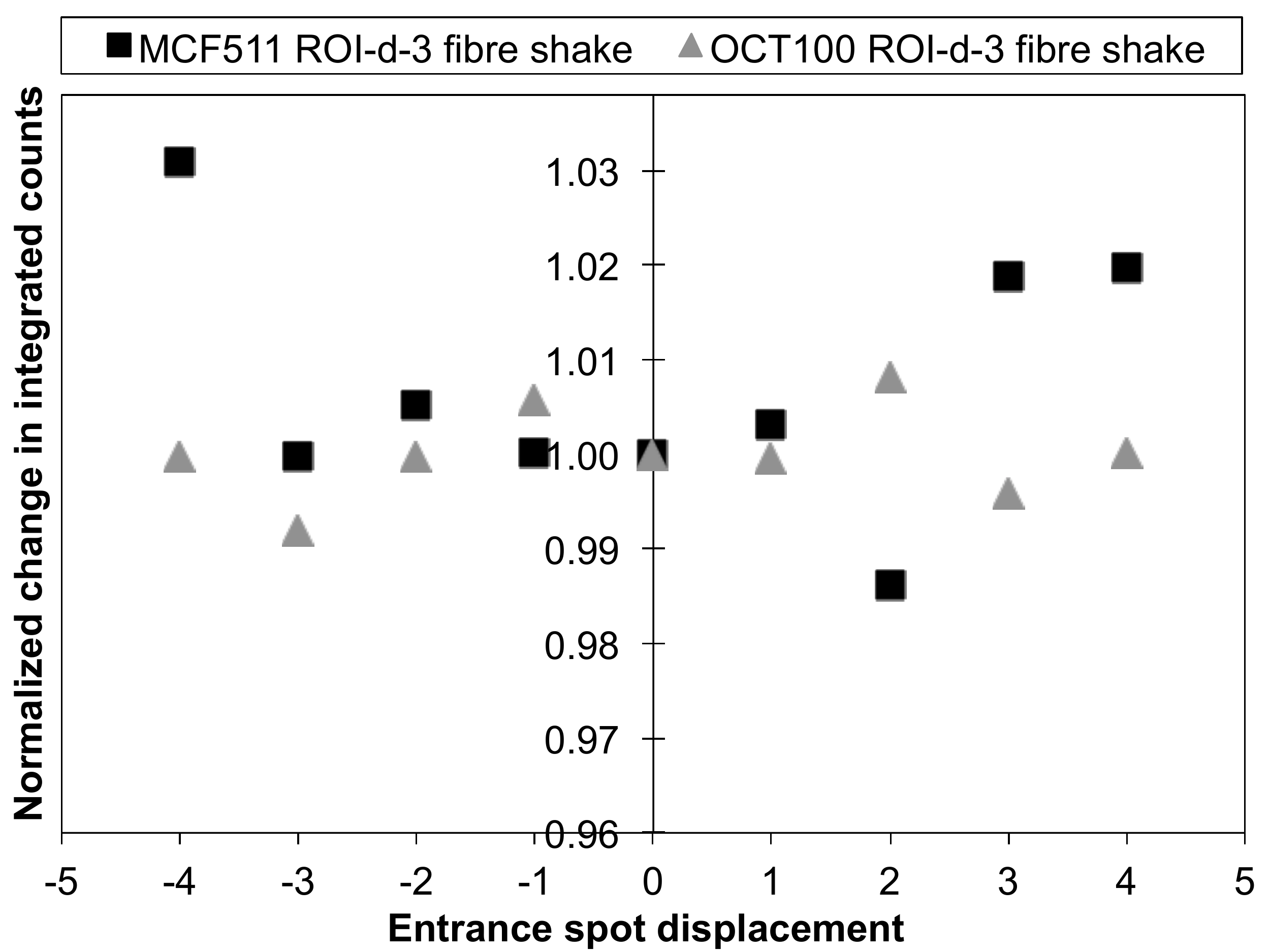}\\
	\includegraphics[width=\columnwidth]{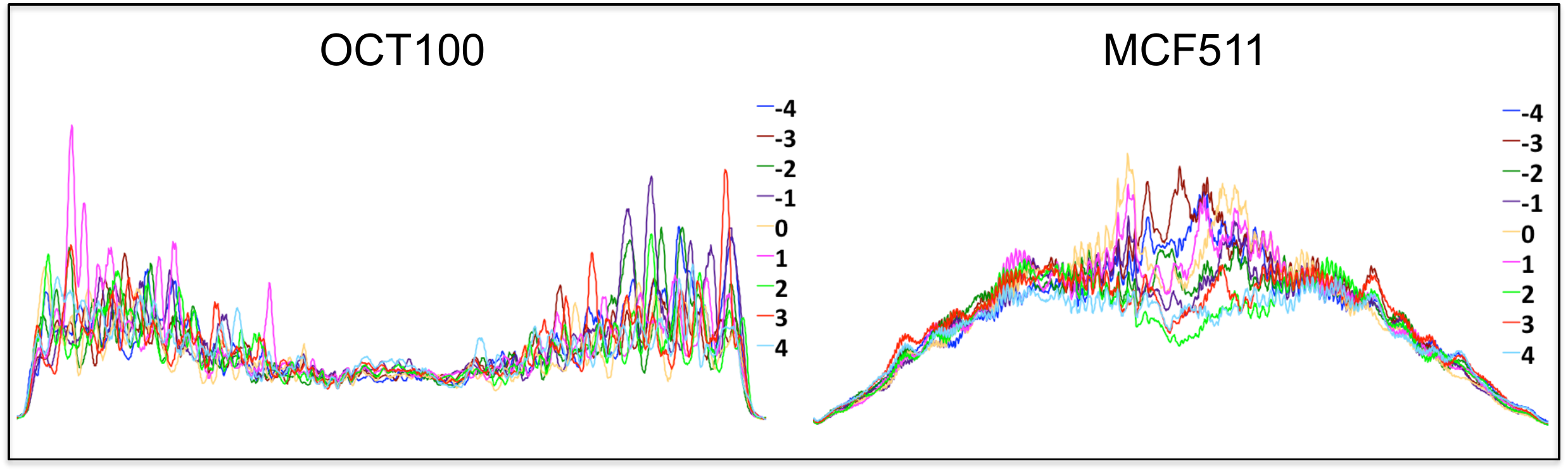}
\end{tabular}
\caption{Top: Far field change in integrated counts (modal noise) resulting from scanning an 8$\mu$m diameter input spot of coherent light at 632.8~nm across the FUT cores with fibre shake and aperture vignetting applied (ROI-d-3). Bottom left: Far field cross section profiles for OCT100 (1-D slice at barycentre for each input spot position). Bottom right: Far field cross section profiles for MCF511 (1-D slices at barycenter for each input spot position).}\label{ff_modalnoise}
\end{figure} 

The change in integrated counts (modal noise) in the far field when aperture vignetting (ROI-d-3) and fibre shake is applied to the FUT is shown in Fig.~\ref{ff_modalnoise} (top). Given that OCT100 supports $\sim$3.3 times as many modes as MCF511 and that one expects a proportionately higher signal-to-noise as a result of lower modal noise due to the greater number of modes propagating \cite{Baudrand2001,Feger2012,Grupp2003,Lemke2011,LeonSaval2010}, it is very promising that the MCF511 performs comparably to the OCT100 fibre. The 1-D profiles shown in Fig.~\ref{ff_modalnoise} (bottom) demonstrate that the MCF511 concentrates the light in the central section with little mode migration, where as OCT100 distributes the light such that a ring like pattern is visible with lower intensity in the central section that is relatively flat and higher intensity in the outer region, implying that the shaking of the OCT100 fibre re-distributes the energy into the higher order modes.

\section{Conclusions}
We have described the principles behind a mode-scrambling relay fibre comprising a multicore fibre with photonic lanterns at each end, with a view to high-spectral-resolution precision radial velocity applications. Externally it behaves as a multimode fibre with the ability to collect more light than a single-mode fibre, though with the advantage (not discussed) of the low bend losses of the single- or few-moded cores in the MCF region. The MCF's different core diameters enhance phase and amplitude scrambling of supermodes, thus suppressing modal noise, while at the same time permitting low-loss operation over a wide range of wavelengths.
 
Despite supporting 3.3 times fewer modes than octagonal OCT100, the multicore MCF511 fibre has comparable performance and should therefore perform significantly better than an equivalently moded octagonal fibre and, in turn, the equivalently moded circular fibre.  Another desirable feature is that as long as the number of modes is matched within the device (\textit{N}$_{MM}$=\textit{N}$_{MCF}$) and with the etendue of the external instrumentation, the device does not cause focal-ratio degradation (FRD). Lower focal-ratio modes are simply not guided. There can only be FRD if the input port is under-filled.

Significant performance improvement of the MCF device should be possible by eliminating the residual core structure issues in the multimode input and output regions, for example by splicing a very short section (<10 mm) of MM fibre onto the PL end, inducing dopant diffusion within the PL end, or fabricating an MCF with the cores spread more evenly throughout the fibre diameter. In addition the team has plans to develop and test an infrared (IR) optimized version to provide efficient amplitude and phase scrambling when fibre mode numbers are relatively low (a few 10's of modes), with the aim to addressing the even-larger modal noise issues that limit the achievable signal-to-noise in fibre fed IR high-resolution spectroscopy \cite{Iuzzolino2014,McCoy2012,Origlia2014}.

\section*{Acknowledgements}

The authors acknowledge the OPTICON Research Infrastructure for Optical/IR Astronomy supported by the European Commission's FP7 Capacities programme (Grant number 312430) and the financial support of the German Federal Ministry of Education, Research (BMBF) Program Unternehmen Region (grant no. 03Z2AN11).


\newcommand{\noopsort}[1]{} \newcommand{\printfirst}[2]{#1}
  \newcommand{\singleletter}[1]{#1} \newcommand{\switchargs}[2]{#2#1}

\end{document}